\documentclass[11pt]{article}
\usepackage{amsmath,amssymb,graphicx,mathpazo,times,fullpage}
\usepackage[colorlinks=true,citecolor=blue]{hyperref}

\makeatletter
\renewcommand\section{\@startsection {section}{1}{\z@}%
  {-2ex \@plus -1ex \@minus -.2ex}%
  {1ex \@plus.1ex}%
  {\normalfont\bf\sffamily}}
\renewcommand\subsection{\@startsection{subsection}{2}{\z@}%
  {-1.75ex\@plus -0.4ex \@minus -.2ex}%
  {0.6ex \@plus .1ex}%
  {\normalfont\small\bf\sffamily}}
\renewcommand\subsubsection{\@startsection{subsubsection}{3}{\z@}%
  {-0.6ex\@plus -0.2ex \@minus -.2ex}%
  {0.4ex \@plus .1ex}%
  {\normalfont\normalsize\it}}
\renewcommand\paragraph{\@startsection{paragraph}{4}{\z@}%
  {0.2ex \@plus0.2ex \@minus0.1ex}{-0.5em}%
  {\normalfont\normalsize\bfseries}}
  {\endlist}
  {\end{list}}
  {\end{list}}
  {\end{list}}
  {\end{list}}

\numberwithin{equation}{section}
\def\ps@headings{%
  \let\@oddfoot\@empty
  \let\@evenfoot\@empty
  \def\@evenhead{\small\sffamily\thepage\hfil\slshape\leftmark}%
  \def\@oddhead{\small\sffamily{\slshape\rightmark}\hfil\thepage}%
      \let\@mkboth\markboth
    \def\chaptermark##1{\markboth{{\ifnum \c@secnumdepth >\m@ne
          \if@mainmatter \@chapapp\ \thechapter. \ \fi \fi ##1}}{}}%
    \def\sectionmark##1{\markright {{\ifnum \c@secnumdepth >\z@
          \thesection. \ \fi ##1}}}}
\makeatother

\advance\textwidth 8mm
\advance\hoffset -4mm
\advance\textheight 2mm
\advance\voffset -1mm
\advance\parskip by 2pt
\def\doibase{http://dx.doi.org/}
\def\reftitle#1{``#1''}
\def\be{\begin{equation}}
\def\ee{\end{equation}}
\def\bse{\begin{subequations}}
\def\ese{\end{subequations}}

\def\maketitle{\par\noindent{\bf\LARGE\sffamily\thetitle}\\[1.4ex]%
{\large\theauthor}\\[0.6ex]%
\textit{\thetextinfo}\\[0.2ex]%
{\small\noindent\today}\par\vglue1.4\bigskipamount}

\def\title#1{\def\thetitle{#1}}

\def\author#1{\def\theauthor{#1}}

\def\textinfo#1{\def\thetextinfo{#1}}

\makeatletter
\def\overl@ss#1#2{\vcenter{\offinterlineskip\ialign{$\m@th#1\hfil##\hfil$\crcr#2\crcr<\crcr } }}
\def\gl{\mathrel{\mathpalette\overl@ss>}}

\def\sech{\mathop{\rm sech}\nolimits}
\def\diag{\mathop{\rm diag}\nolimits}

\def\dn{\mathop{\rm dn}\nolimits}
\def\Real{\mathbb{R}}
\def\Complex{\mathbb{C}}
\def\Integer{\mathbb{Z}}
\def\Re{\mathop{\rm Re}\nolimits}
\def\Im{\mathop{\rm Im}\nolimits}

\def\tr{\mathop{\rm tr}\nolimits}

\def\d{\mathrm{d}}
\def\sgn{\mathop{\rm sgn}\nolimits}
\def\e{\mathop{\rm e}\nolimits}
\def\@#1{{\mathbf{#1}}}
\def\_#1{{\mathsf{#1}}}
\let\~=\tilde
\let\==\bar
\let\^=\hat
\let\<=\langle
\let\>=\rangle

\def\Leps{\mathop{\mathcal{L}^\epsilon}\nolimits}

\def\min{\text{min}}
\def\max{\text{max}}
\def\mf#1{\mathcal{#1}}
\def\tp{p}
\makeatother

\let\uppercase=\relax

\let\bb=\relax
\let\eb=\relax
\def\bm{\begingroup\blue}
\def\em{\endgroup}
\let\bm=\relax
\let\em=\relax

\let\eref=\eqref
\advance\textwidth -12mm
\advance\hoffset 6mm
\advance\textheight 4mm
\advance\voffset -2mm

\advance\parskip 0.4pt

\let\trueparagraph=\paragraph
\def\paragraph#1{\par\smallskip\trueparagraph{\rm\textbf{#1}}}

\newdimen\figwidth
\allowdisplaybreaks

\begin{document}

\title{Semiclassical dynamics and coherent soliton condensates in self-focusing nonlinear media with periodic initial conditions}
\author{Gino Biondini$^{1,2}$ and Jeffrey Oregero$^1$}

\textinfo{$^1$ Department of Mathematics, State University of New York, Buffalo, NY 14260\\
$^2$ Department of Physics, State University of New York, Buffalo, NY 14260}

\maketitle


\noindent\textbf{Abstract.}
The small dispersion limit of the focusing nonlinear Schr\"odinger equation with periodic initial conditions
is studied analytically and numerically. 
First, through a comprehensive set of numerical simulations, 
it is demonstrated that solutions arising from a certain class of initial conditions, referred to as ``periodic single-lobe'' potentials,
share the same qualitative features, 
which also coincide with those of solutions arising from localized initial conditions.
The spectrum of the associated scattering problem in each of these cases is then numerically computed, 
and it is shown that such spectrum is confined to the real and imaginary 
axes of the spectral variable in the semiclassical limit.
This implies that all nonlinear excitations emerging from the input have zero velocity, 
and form a coherent nonlinear condensate.
Finally, by employing a \bm formal \em Wentzel-Kramers-Brillouin expansion for the scattering eigenfunctions,
asymptotic expressions for the number and location of the bands and gaps in the spectrum are obtained,
as well as corresponding expressions for the relative band widths
and the number of ``effective solitons''.
These results
are shown to be in excellent agreement with those from direct numerical computation of the eigenfunctions.
In particular, a law is obtained describing how the number of effective solitons scales with the
small dispersion parameter.

\noindent\textit{To appear in Studies in Applied Mathematics}


\bigskip

\section{\uppercase{Introduction}}

Many physical systems are characterized by the simultaneous presence of dispersion and nonlinearity.
The combination of these two effects can produce a wide variety of physical phenomena,
ranging from modulational instability, collapse and supercontinuum generation to the formation of solitons, rogue waves, dispersive shocks, 
wave turbulence etc.
(e.g., see \cite{solli,zakharov2009,DudleyTaylor,randoux2014,az2014,onoratoosborne,elhoefer,Whitham}
and references therein).

Often, the typical scales in the system are such that nonlinear effects are much stronger than dispersive ones.
These kinds of problems are referred to as small-dispersion (or semiclassical) limits.
The canonical example is perhaps that of the Korteweg-deVries (KdV) equation.
Indeed, it was the desire to understand the Fermi-Pasta-Ulam recurrences
via the behavior of solutions in the small-dispersion limit of the KdV equation
that led to the discovery of solitons in the first place \cite{ZabuskyKruskal}
as well as to the development of the inverse scattering transform (IST) to solve the initial value problem for the KdV equation \cite{GGKM1967}.
The IST was then used to study the small-dispersion limit of the KdV equation analytically in \cite{LaxLevermore}
and many works thereafter.

While the KdV equation provided the initial impetus for these discoveries,
many nonlinear dispersive systems are governed by the nonlinear Schr\"odinger (NLS) equation.
Indeed, the NLS equation is known to be a universal model for the evolution of nonlinear dispersive wave trains \cite{BN1967,CalogeroEckhaus}.
As such, it arises in such diverse fields as water waves, plasmas, optics and Bose-Einstein condensates \cite{AS1981,Agrawal2007,IR2000,PS2003,KA2003,KFCG2008}.
Like the KdV equation, the NLS equation is also a completely integrable Hamiltonian system, and as a result a number of analytical techniques 
such as the IST are available to study the behavior of its solutions \cite{AS1981,ZS1972,NMPZ1984,FT1987,TrogdonOlver}.
The NLS equation comes in two variants: 
the defocusing case (arising with normal dispersion in optical fibers and repulsive Bose-Einstein condensates) 
and the focusing case (arising in water waves, anomalous dispersion, and attractive condensates).
Typically, the dispersive and nonlinear effects in the NLS equation should be comparable in order to obtain solitons.  
However, in many physical scenarios the nonlinearity is much stronger than dispersion.  
For example, this happens with high-power input lasers or high-nonlinearity fibers in optics.  
These regimes give rise to strongly nonlinear phenomena.  
In previous works we showed that, for the KdV and defocusing NLS equations, 
in many cases the resulting dynamics is characterized by the generation of a large number of ``effective solitons''.  
In the present work we show that the same is true in the focusing case.

The semiclassical limit of the focusing NLS equation has been studied extensively 
\cite{millerkamvissis1998,bronskikutz1999,KMM2003,lyng2012,TVZ2004,ClarkeMiller2002,
JM2013,EKT2016,bertolatovbis}.
%
Previous works however considered localized initial conditions (ICs). 
For 
the defocusing 
NLS equation, 
the thermodynamic limit of solutions generated by a special class of ICs with non-zero background was studied in~\cite{Conti09}. 
In the defocusing case with periodic ICs, 
the small dispersion limit was recently realized in fiber optics experiments, which show fission of dark solitons from periodic breaking points 
\cite{trillovaliani,fatometrillo}.
These results were then characterized analytically in \cite{pre2017deng}.
Experimental studies on related nonlinear problems were also recently reported in 
\cite{wetzel2016,bongiovanni2019,podivilov2019,kraych2019}.
\bb
It should be mentioned that observing the semiclassical regime of the focusing NLS equation experimentally involves a very delicate and careful set-up,
since small values of the semiclassical parameter $\epsilon$ in Eq.~\eqref{e:NLS} below imply that any higher-order physical effects present in the system 
might spoil the phenomena that one is seeking to observe.
Fiber optic experiments were reported that are equivalent to values of $\epsilon$ as small as 0.002
\cite{sudo}.
\bm
It was also recently shown experimentally that the semiclassical description of fNLS is still valid for not so small values of $\epsilon$ as well,
see for example \cite{tikan}.
\em 
Nonetheless, the experimental observation of detailed semiclassical behavior in the anomalous dispersion regime in fiber optics \bm is still challenging problem 
\cite{boscolofinot}. \em
Moreover, 
to the best of our knowledge, 
\eb
no analytical studies are available on the behavior of solutions of the semiclassical focusing NLS equation 
with periodic ICs.

In this work we report an analytical and numerical study of focusing \bm periodic \em dispersive media
in a strongly nonlinear regime.
First, 
through a comprehensive set of numerical simulations, 
we show in Section~\ref{s:nls}
that solutions arising from many different initial conditions, referred to as ``periodic single-lobe'' potentials, 
share the same qualitative features, 
which coincide with those of solutions arising from localized ICs. 
Then in Section~\ref{s:spectrum} we
compute the spectrum of the associated scattering problem, 
and we show that the spectrum is entirely confined to the real and imaginary 
axes of the spectral variable in the semiclassical limit.
This implies that all nonlinear excitations emerging from the input have zero velocity, 
and form a coherent nonlinear condensate.
Finally, 
in Section~\ref{s:wkb},
by employing a \bm formal \em Wentzel-Kramers-Brillouin (WKB) expansion for the scattering eigenfunctions,
we obtain asymptotic expressions for the number and location of the bands and gaps in the spectrum,
as well as corresponding expressions for the relative band widths,
which are in excellent agreement with direct numerical computation of the eigenfunctions.
In particular, we show that the problem naturally leads one to formulate the concept of ``effective solitons'',
and we obtain a law describing the scaling of the number of effective solitons 
as a function of the
small dispersion parameter.
\bm
Section~\ref{s:numerics} provides 
a discussion of the various numerical methods used, further numerical results,
while 
section~\ref{s:wkbdetails} provides some details of the WKB calculations.
\em
We conclude this work with a discussion and some final remarks in Section~\ref{s:discussion}.

\section{\uppercase{Semiclassical focusing NLS equation with single-lobe periodic potentials}}
\label{s:nls}
The starting point for our study is the focusing NLS equation in the semiclassical regime, 
namely 
\be
i\epsilon q_t + \epsilon^2 q_{xx} + 2|q|^2 q = 0\,,
\label{e:NLS}
\ee
where 
$q(x,t)$ 
is the slowly varying complex envelope of a quasi-mono\-chromatic, weakly dispersive nonlinear wave packet,
subscripts $x$ and $t$ denote partial derivatives
and the physical meaning of the variables $x$ and $t$ depends on the physical context. 
(E.g., in optics, $t$ represents propagation distance while $x$ is a retarded time.) 
The parameter $\epsilon$ quantifies the relative strength of dispersion compared to nonlinearity.
(In quantum-mechanical settings, $\epsilon$ is also proportional to Planck's constant $\hbar$.)
Of course, both instances of $\epsilon$ in Eq.~\eqref{e:NLS} could be scaled away via suitable changes of independent and dependent variables.
However, the solutions of Eq.~\eqref{e:NLS} also depend on the ICs, and the corresponding transformations 
would produce ICs that depend on~$\epsilon$.
In other words, studying the semiclassical limit corresponds to the study of the behavior of solutions of Eq.~\eqref{e:NLS}
with fixed ICs as $\epsilon\downarrow0$.

\subsection{Initial conditions}

Here we study the dynamics of solutions of Eq.~\eqref{e:NLS} generated by a certain class of ICs which we refer to as ``single-lobe periodic potentials''. 
Specifically, 
we call a single-lobe periodic potential 
the continuous periodic extension of a 
real-valued function $q: [-L,L] \to \Real$ 
\bb
for which 
(i) $q(-L) = q(L)$ and (ii)
\eb
there exists a point $x_{\max} \in (-L,L) $ such that $q(x)$ is increasing on $(-L, x_{\max})$ and decreasing on $(x_{\max}, L)$.
(Here we used the translation invariance of the NLS equation and the corresponding Zakharov-Shabat scattering problem so that the minimum of 
the potential is obtained at $x=\pm L$.)\,\ 
To the best of our knowledge, potentials of this form had only been studied on the infinite line 
\cite{millerkamvissis1998,bronskikutz1999,KMM2003,lyng2012,KS2002,KS2003,BL2018}.  
Moreover, 
for simplicity in all the examples discussed in this work we also assume that 
$q(x)$ is even and 
$q(\pm L) \ge$ 0.
These last two conditions will simplify the calculations of the asymptotic behavior of the spectrum.

In particular, we will consider the following specific examples of single-lobe periodic ICs as distinguished cases:
\vspace*{-0.4ex}
\bse
\label{e:IC}
\begin{gather}
q_\mathrm{cos}(x,0) = (1+\cos x)/2\,,
\label{e:IC1}
\\
q_\mathrm{expsin}(x,0) = \e^{-\sin^2x}\,,
\label{e:IC2}
\\
q_\mathrm{dn}(x,0) = \dn(x|m)\,, \; 0<m<1.
\label{e:IC3}
\end{gather}
\ese
The shape in Eq.~\eqref{e:IC1}, commonly referred to as a ``raised cosine'', 
is easily generated experimentally 
\bb and \eb
is quite common in optical communications
\cite{Gowar1993,Agrawal2002}.
Here and below, $\dn(\cdot|m)$ is one of the 
Jacobian elliptic functions, and~$m$ the corresponding elliptic parameter \cite{NIST}.
Recall that 
$\dn(x|0)=1$ while $\dn(x|1)= \sech\,x$.
Hence, when $m=1$ the problem reduces to that studied in \cite{millerkamvissis1998,bronskikutz1999,KMM2003,lyng2012}.
More in general, the real period of Eq.~\eqref{e:IC3}
is $2K(m)$, where $K(\cdot)$ is the complete elliptic integral of the first kind~\cite{NIST}.
One of the main points of this work, however, is that the dynamics are relatively insensitive to the specific input, 
and many different choices of ICs would lead to similar results.
See further discussion in section~\ref{s:numerics}.

\subsection{Dynamical behavior}

\begin{figure}[t!]
\kern1ex
\centerline{\includegraphics[width=7.5cm]{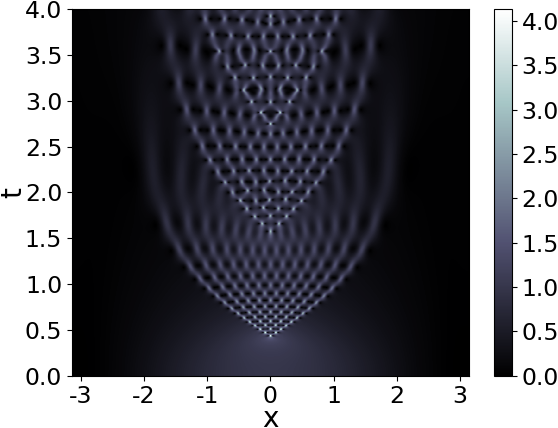}
\hspace{4mm} 
\includegraphics[width=7.5cm]{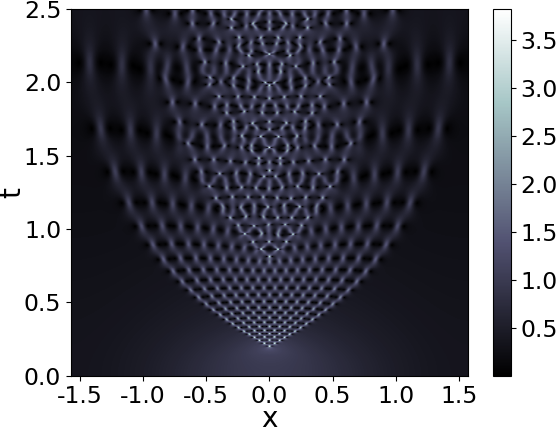}}
\vspace{2.5mm}
\centerline{\includegraphics[width=7.5cm]{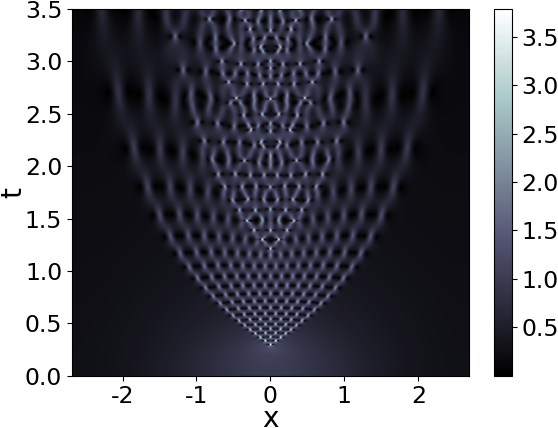}
\hspace{4mm}
\includegraphics[width=7.5cm]{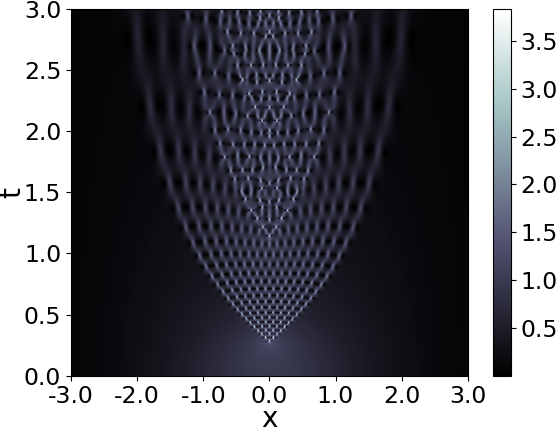}}
%
\caption{Density plot of the amplitude $|q(x,t)|$ of the solution of the focusing NLS equation in the semiclassical limit with different ICs.
The horizontal axis is the spatial variable $x$ and the vertical axis is time $t$.
Top left: the ``raised cosine'' IC in Eq.~\eqref{e:IC1} with $\epsilon = 0.06$.
Top right: the exp-sine IC in Eq.~\eqref{e:IC2} with $\epsilon = 0.026$.
Bottom left: the dn IC in Eq.~\eqref{e:IC3} with $m=0.92$ and $\epsilon = 0.044$.
For comparison purposes, the bottom right panel shows a density plot of the numerical solution of the focusing NLS equation with localized IC
$q(x,0) = \text{sech} x$ and $\epsilon = 0.037$.
See section~\ref{s:numerics} for a demonstration of the dynamical behavior as $\epsilon \downarrow 0$.}
\label{f:1}
\end{figure}

We numerically integrated Eq.~\eqref{e:NLS} with IC given by Eq.~\eqref{e:IC} using an eighth-order Fourier split-step method \cite{yoshida, tappert, fornberg, yang, weideman} in double precision.
All results were checked for numerical convergence (see section~\ref{s:numerics} for further details).
Figure~\ref{f:1} shows density plots of the numerically computed amplitude $|q(x,t)|$ using 
the raised cosine IC \eqref{e:IC1} with $\epsilon = 0.06$ (top left), 
the exp-sine IC \eqref{e:IC2} with $\epsilon = 0.026$ (top right), and 
the dn IC \eqref{e:IC3} with $m=0.92$ and $\epsilon = 0.044$ (bottom left).  
For comparison we also include an IC on the infinite line, namely, $q(x,0) = \sech x$ with $\epsilon = 0.037$ (bottom right).

It is well known that, 
in the focusing NLS equation on the line (i.e., for localized ICs as in the bottom right panel of Fig.~\ref{f:1}), 
\bm
the focusing dynamics often
(but not necessarily) 
results in a focusing singularity or a gradient catastrophe.
(The gradient catastrophe typically occurs in both the amplitude and the phase of the solution.
Note however that the chirp, either from the ICs or developed in the process of the time evolution, 
can either accelerate or slow down, or even prevent, the occurrence of the gradient catastrophe.)
More specifically, what one sees in Fig.~\ref{f:1} is a typical picture of primary gradient catastrophe of a modulated plane wave, followed by similar catastrophes of higher genus solutions.
\em
The singularity is regularized by the weak dispersion, and the subsequent generation of a 
complex oscillation structure corresponding to a slow modulation of the genus-2 solutions of the focusing NLS equation
\cite{KMM2003,lyng2012,TVZ2004,EKT2016,bertolatovbis}.
A secondary breaking is also present, beyond which the asymptotic analysis of the inverse problem in the IST breaks down \cite{KMM2003,lyngmiller}. 
Numerical evidence and the asymptotics of the inverse problem in the IST both suggest that, after the secondary breaking, the solution is described by a slow modulation of genus-4 solutions
\cite{EKT2016, lyngmiller},
and the possible existence of further breakings was also conjectured there.
(Indeed, Fig.~\ref{f:1} clearly indicates the presence of a tertiary breaking beyond which one might have genus-6 behavior.)
\bb
The spatial and \bm temporal \em period of the small-scale oscillations is proportional to $\epsilon$, 
and therefore the limit $\epsilon\downarrow 0$ can only be interpreted in a weak sense. 
Nonetheless, the large-scale structure of oscillations (and in particular the breaking time and the location of the caustic curves) become independent of~$\epsilon$ in the dispersionless limit.
This phenomenon is also observed with periodic boundary conditions, as illustrated in Fig.~\ref{f:S1a} of section~\ref{s:numerics} for a specific choice of potential, namely Eq.~\eqref{e:IC2}. 
\eb

\bb
Most importantly, however, the results shown in Fig.~\ref{f:1} clearly demonstrate that the semiclassical behavior of solutions to the focusing NLS equation on the infinite line ---
namely a sequence of three breakings each leading to the formation of higher-genus oscillations --- is also observed with periodic boundary conditions. 
In other words,
\eb
Fig.~\ref{f:1} demonstrates that the behavior of solutions of the focusing NLS equation in the semiclassical limit 
displays universal features, independently of the ICs and of whether such ICs are periodic or localized.
(Of course one should not interpret the above statement as saying that \textit{all} ICs give rise to this behavior,
and other scenarios are also possible;
see the discussion in sections~\ref{s:numerics} and~\ref{s:discussion} for further details.
\bm
We also note that the universality of the first gradient catastrophe was proved in
\cite{bertolatovbis}.\em)

The above result is in marked constrast to the semiclassical limit of the Korteweg-deVries (KdV) and of the defocusing NLS equations,
where the dynamics results in the formation of solitons that separate from each other and travel independently.
It was argued in~\cite{dubrovin}, 
\bm 
numerically investigated in~\cite{dubrovingravaklein},
and proved in~\cite{bertolatovbis} in specific situations
\em 
that the behavior of solutions near the first breaking point 
(i.e., the gradient catastrophe) possesses universal features, 
which for the focusing NLS equation are described in terms of the Tritonqu\'ee solution of the Painlev\'e I equation.
A precise asymptotic characterization of the oscillation pattern after the the first breaking was also obtained in \cite{bertolatovbis},
and is also described by the Tritronqu\'ee solution.
All these analytical results, however, as well as those mentioned in the previous paragraph, are limited to the NLS equation with localized IC.
Indeed, Fig.~\ref{f:1} shows that within the class of single-lobe potentials the qualitative features of the solution are the same, 
independently of whether the ICs are periodic or localized and also independently of the specific details of the ICs. 


\section{\uppercase{NLS spectrum in the semiclassical limit}}
\label{s:spectrum}

Some of the features discussed above can be characterized analytically by taking advantage of the mathematical tools associated with 
the complete integrability of the NLS equation.

\subsection{Lax pair and monodromy matrix}
Recall that Eq.~\eref{e:NLS} is the compatibility condition of the matrix Lax pair \cite{ZS1972}
\vspace*{-0.6ex}
\bse
\label{e:LP}
\begin{gather}
\epsilon \phi_x = X\phi\,,
\label{e:LP1}
\\
\epsilon \phi_t = T\phi\,,
\end{gather}
\ese
\bm 
where 
$\phi(x,t,\zeta)$ is a simultaneous solution of both parts of~\eqref{e:LP}, with
\em
\vspace*{-0.8ex}
\bse
\begin{gather}
\hspace{-3mm} 
X(x,t,\zeta)= -i\zeta \sigma_3 + Q\,,
\\
T(x,t,\zeta)= -i(2\zeta^2 + |q|^2 + \epsilon Q_x)\,\sigma_3 + \zeta Q\,,
\label{e:SP}
\end{gather}
\ese
where 
$\sigma_3 = \diag(1,-1)$ is the third Pauli matrix,  and
\vspace*{-1ex}
\be
Q(x,t) = \begin{pmatrix} 0 &q(x,t)\\ - q^*(x,t)& 0 \end{pmatrix}.
\ee
The first half of the Lax pair [i.e., Eq.~\eqref{e:LP1}], $\zeta$ and $q(x,t)$ are referred to as the Zakharov-Shabat (ZS) scattering problem, 
scattering parameter and scattering potential, respectively.
Equation~(\ref{e:LP1}) can also be written as the eigenvalue problem
\vspace*{-1ex}
\be
\Leps \phi = \zeta \phi\,,
\label{e:eigval}
\ee
where $\Leps$ is the one-dimensional Dirac operator 
\be
\Leps = i\sigma_{3}( \epsilon \partial_{x} - Q)\,.
\label{e:op}
\ee
Thus, $\zeta$ and $\phi(x,t,\zeta)$ are also referred to as the eigenvalue and the corresponding eigenfunction, respectively.
The Lax spectrum $\Sigma(\Leps)$ of $\Leps$ is the set of all values of $\zeta\in\Complex$ for which nontrivial bounded solutions $\phi(x,t , \zeta)$ 
of Eqs.~\eqref{e:LP} exist. 

The inverse scattering transform (IST) allows one to solve the initial-value problem for Eq.~\eqref{e:NLS} 
by associating to $q(x,t)$ suitable scattering data via the solutions of the scattering problem.
Once the scattering data are obtained from the initial condition,
$q(x,t)$ is reconstructed in terms of the scattering data by inverting the scattering transform
\cite{AS1981,NMPZ1984,APT2004}.

Floquet-Bloch theory \cite{Floquet,MW1966,Eastham} implies that, when the potential in Eq.~\eqref{e:LP1} is $2L$-periodic,
all bounded solutions are of the form 
\vspace*{-1ex}
\be
\phi(x,\zeta) = \e^{i\nu x}w(x,\zeta)\,,
\label{e:floq}
\ee
where $w(x+2L,\zeta) = w(x,\zeta)$, $ i\nu $ is referred to as the Floquet exponent,
$\nu \in [0, \pi/L) $,
and the time dependence was omitted for brevity.
Moreover, the Floquet multipliers $\mu = \e^{2i\nu L}$ are the eigenvalues of the monodromy matrix $M(\zeta)$, 
defined as 
\vspace*{-1ex}
\be
M(\zeta) = \Phi(x-L,\zeta)^{-1}\Phi(x+L,\zeta)\,,
\label{e:Mdef}
\ee
where $\Phi(x,\zeta)$ is any fundamental matrix solution of Eq.~\eqref{e:LP1}. 
Since $\text{det}M \equiv 1$, 
the eigenvalues of $M$ are the roots of the polynomial $\mu^2 - (\tr M)\,\mu + 1 = 0$, and
it follows that Eq.~\eqref{e:LP1} has bounded solutions 
if and only if $\zeta$ is such that 
$\tr M \in \Real$ and 
$-2\le \tr M \le 2$. 
The Floquet-Bloch spectrum of $\Leps$ is then given by
\be
\Sigma_\nu(\Leps) = \{\zeta \in \Complex : \tr M(\zeta) = 2\cos(2\nu L) \}\,,
\label{e:Floquetspec}
\ee
and 
the Lax spectrum is the union of all Floquet-Bloch spectra:
$\Sigma(\Leps) = \cup_{\nu\in[0,\pi/L)}\Sigma_\nu(\Leps)$.
The NLS equation~\eqref{e:NLS} amounts to an isospectral deformation of $\Leps$; therefore, $\tr M(\zeta)$, 
$\Sigma_\nu(\Leps)$ and $\Sigma(\Leps)$ are independent of time.
However, $\Leps$ is non-self-adjoint, which complicates the problem significantly, since it means that the spectrum is in general complex.
Nonetheless, the symmetries of the scattering problem imply that 
the Lax spectrum is always symmetric with respect to the real $\zeta$-axis.
Moreover, if $q(x,t)$ is even with respect to $x$, the spectrum is also symmetric with respect to the imaginary $\zeta$-axis.

\subsection{Numerical evaluation of the Lax spectrum}

We next show that the Lax spectrum of the ZS operator simplifies considerably in the semiclassical limit. 
Recall that the focusing ZS scattering problem on the line
[i.e., with potentials $q\in L^1(\Real)$]
posseses both a continuous and a discrete spectrum, 
with the former consisting of the real $\zeta$-axis, 
whereas the latter can be fairly complicated 
\cite{zhou,deiftzhou},
\bb
even though for single-lobe potentials the discrete spectrum is confined to the imaginary $\zeta$-axis \cite{KS2002,KS2003}.
\eb
The semiclassical limit of the ZS problem was studied numerically in~\cite{bronski}, 
and formal WKB calculations were reported in~\cite{KMM2003,miller_physd2001}, 
while an unpublished result by Deift, Venakides and Zhou 
states that, as $\epsilon\downarrow 0$, 
the discrete eigenvalues of the ZS problem on the line with real-valued potentials accumulate to the real and imaginary axes of the spectral plane.
(A modified version of their result can be found in section~3 of \cite{difrancomiller}.)
All of the above results, however, apply to potentials on the infinite line, not to periodic potentials.

For periodic potentials, 
the Lax spectrum of the ZS problem is composed of a (possibly infinite) number of spectral bands,
\bb
each spectral band consisting of a (finite or infinite) curve along which 
$\tr M(\zeta) \in [-2, 2]$
[cf.~\eqref{e:Floquetspec}].
\eb
\bm
Since $q$ is $2L$-periodic \em
the band edges correspond to the Floquet-Bloch spectrum for \bm $\nu = \pi/L$,
and $\nu = \pi/2L$\em,
which in turn is associated with periodic and anti-periodic eigenfunctions, respectively.
It was recently proved in \cite{fujiiewittsten2018} that 
the periodic eigenvalues, 
\bm i.e., the Floquet-Bloch spectrum with $\nu=n\pi/L, n\in\Integer$ \em
of Eq.~\eqref{e:LP1} 
with real-analytic periodic potentials concentrate on the real and imaginary $\zeta$-axes as $\epsilon \downarrow  0$. 
This is a powerful result, which applies to general real-analytic periodic potentials (i.e., not only single-lobe).
On the other hand, it does not provide any information about the Floquet-Bloch spectrum for \bm $\nu\ne n\pi/L$ \em.
In practice, this means that, even though half of the band edges converge to the real and imaginary axis,
no information is available about the behavior of the full spectral bands.
To investigate this question, we therefore turn to numerics.

\begin{figure}[t!]
\centerline{\includegraphics[width=7.5cm]{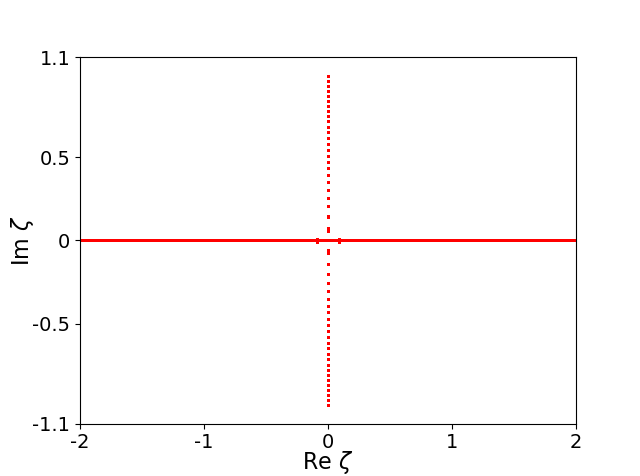}
\includegraphics[width=7.5cm]{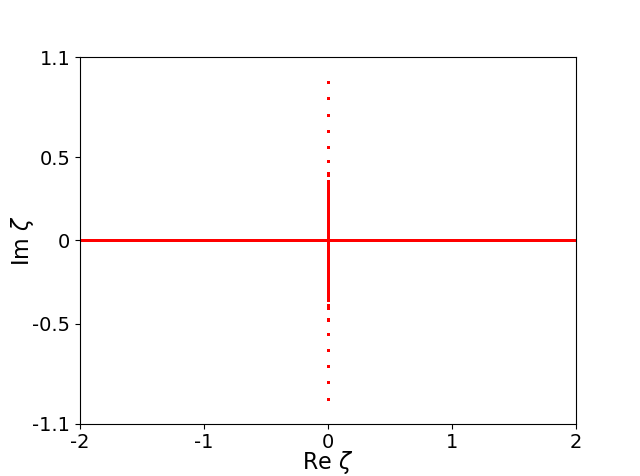}}
\centerline{\includegraphics[width=7.5cm]{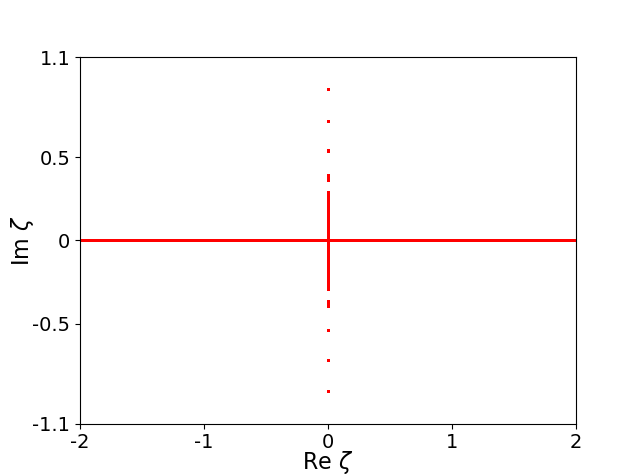}
\includegraphics[width=7.5cm]{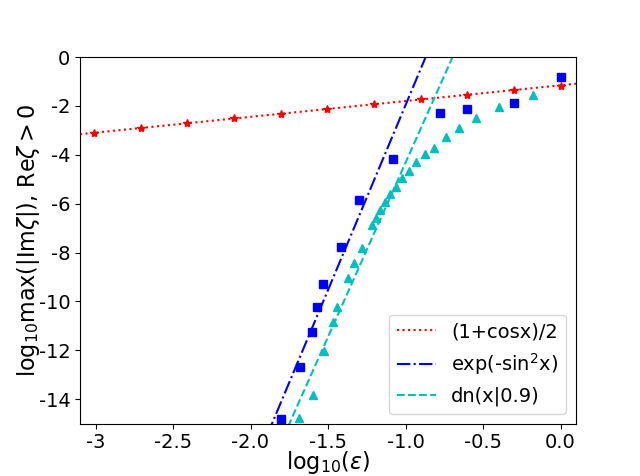}}
%
\caption{The spectrum (red bands) of the scattering problem Eq.~\eqref{e:LP1} as computed numerically via Floquet-Hill's method. \bb
Top left: the raised cosine potential in Eq.~\eqref{e:IC1} with $\epsilon = 0.04$. 
Top right: the exp-sine potential in Eq.~\eqref{e:IC2} with $\epsilon = 0.07$. Bottom left: the dn IC in  Eq.~\eqref{e:IC3} with $m=0.9$ and $\epsilon=0.2$. Bottom right: Convergence of non-imaginary eigenvalues to the real axis as $\epsilon\downarrow0$. 
The stars, triangles and squares are numerically computed data points;
the dotted, dashed and dot-dashed lines are a linear regression fit. \eb
}
\label{f:2}
\end{figure}

Using Floquet-Hill's method \cite{DK2006}, we performed a series of careful numerical simulations of the focusing ZS problem in the semiclassical limit with periodic potentials.
The results, some of which are shown in Fig.~\ref{f:2} (see section~\ref{s:numerics} for further details), reveal
persistent features of the Lax spectrum across a varierty of single-lobe periodic potentials, namely: 
(i) an infinitely long band along the real $\zeta$-axis, 
\bb
as well as a continuous band along the segment $[-iq_\min,iq_\min]$ of the imaginary $\zeta$-axis,
\unskip
\eb
\break
(ii) the absence of any spectral bands in the strips $|\Im\zeta| > |q|_{\max}$, and
(iii) a sequence of bands and gaps on the interval $(iq_{\min}, iq_{\max})$ of the imaginary $\zeta$-axis. 
Most interestingly, however, the numerical evidence strongly suggests that the Lax spectrum in the semiclassical limit is confined to the real and imaginary axes. 
Indeed, a numerical convergence study (see the bottom right panel of Fig.~\ref{f:2}) shows that, 
for eigenvalues off the imaginary axis (i.e., for $\Re\zeta\ne0$), 
one has $\max(|\Im \zeta|) = O(\epsilon^\alpha)$ as $\epsilon \downarrow  0$, with 
\bb
$\alpha = 0.65 \pm 0.013$ for Eq.~\eqref{e:IC1},
$\alpha = 15.1\pm 4.27$ for Eq.~\eqref{e:IC2},
and $\alpha = 14.2 \pm 1.89$ for Eq.~\eqref{e:IC3}, where
\eb
the intervals represent 99\% confidence bands about the slope of the linear regression fit. 
These results are also confirmed by directly computing the scattering eigenfunctions via numerical integration of Eq.~\eqref{e:LP1} 
and using the results to construct the monodromy matrix. 
Of course the spectra arising from different choices of potentials are quantitatively different.
On the other hand, we find it remarkable that all of them display the same qualitative features.
In fact, the properties of the Lax spectrum may be more general, and hold for a large class of \bb real (complex) \eb potentials.

The fact that the spectrum is confined to the real and imaginary axes in the semiclassical limit has an important practical consequence.
Recall that, for the focusing nonlinear Schrodinger equation (NLS) on the infinite line:
(i) each discrete eigenvalue generates a soliton, and
(ii) the real part of the eigenvalue is \bb proportional to \eb the soliton speed.
This means that, if all discrete eigenvalues lie on the imaginary axis, 
all the solitons will have zero velocity, and will therefore generate a bound state.
Some such situations were recently studied in~\cite{LiBSchiebold}.
Moreover these soliton bound states become increasingly complex as the number of solitons increase.

The situation is more complicated in the periodic case, since here one never has true solitons, and must deal with more complex nonlinear excitations instead.
Nonetheless, a similar result emerges, namely that the velocity of these nonlinear excitations is proportional to the real part of the corresponding eigenvalues
\cite{kamchatnov}.
Thus, the above results already have an important practical consequence, since they demonstrate that, in the small dispersion limit, 
the focusing NLS dynamics is very different to that for the KdV and defocusing NLS equations. 
There, each soliton has a different velocity, and therefore they all fly away from each other.  
In contrast, here all the solitons have zero velocity, and the solution is characterized by a coherent soliton condensate,
as we discuss in detail next.

\section{\uppercase{Semiclassical soliton condensates}}
\label{s:wkb}

Next we analyze \bb in more detail \eb the properties of the spectrum and the resulting NLS dynamics
in the semiclassical limit.  
Since the spectrum is independent of time, for brevity we will omit the time dependence in the potential $q$ and the eigenfunctions $\phi$.

\subsection{Asymptotic analysis of the scattering problem}

The invertible change of variables $v = \phi_{1} + i \phi_{2} $ and $ \tilde{v} = \phi_{1} - i\phi_{2}$ 
maps Eq.~\eqref{e:LP1} into the time-independent Schr\"odinger equation with a complex potential, namely
\vspace*{-0.6ex}
\be
\epsilon^{2} v'' + (i\epsilon q'(x) + Z(x,\lambda))\,v = 0\,,
\label{e:ode}
\ee
\vspace*{-0.6ex}
where for convenience we defined 
\be
Z(x, \lambda) = \lambda + q^2(x),
\label{e:Z}
\ee
with $\lambda = \zeta^2$. 
This formulation immediately suggests the use of the WKB method to obtain an asymptotic description of the Lax spectrum. 
In our case, however the situation is complicated by the fact that the spectral problem in Eq.~\eqref{e:ode} is non-self-adjoint, 
and the 
\bm 
use of the WKB method in such situations is known to be challenging 
(cf.\ ``WKB paradox'' in \cite{bronski}). 
\em
%
We note, however, that even though the eigenfunctions $v(x,\lambda)$ are rapidly varying in $\epsilon$ 
(due to the coefficient $\epsilon^2$ in front of the second derivative),
$q(x)$ is independent of $\epsilon$, and therefore the term $i\epsilon q'(x)$ is expected to be a higher-order contribution.
In other words, Eq.~\eqref{e:ode} is formally a small perturbation of Hill's equation \cite{MW1966}.
(Indeed, it was already remarked in~\cite{ZS1972} that the focusing Zakharov-Shabat scattering problem becomes formally self-adjoint in the semiclassical limit.)
This observation, and the strong numerical evidence presented earlier, both suggest that, despite the fact that Eq.~\eqref{e:ode} is not a self-adjoint problem,
the WKB method can still be effective in describing the asymptotic properties of the spectrum in the semiclassical limit.
We next show that this is indeed the case.

For brevity we limit ourselves to reporting the results of our \bm formal \em WKB analysis, omitting the details of the calculations (see section~\ref{s:wkbdetails} for further details).
When $q(x)$ in Eq.~\eqref{e:LP1} is a single-lobe periodic potential, 
the real $\lambda$-axis divides into three disjoint regions,
depending on the possible existence of turning points,
i.e., values of $x$ at which $Z(x,\lambda)=0$.
More precisely:

(i) 
For $ \lambda \in (-\infty, -q^{2}_\max)$, one has $Z(x,\lambda)<0$ for any $x\in[-L,L]$.  
Hence there are no turning points, and the WKB expansion immediately yields
\vspace*{-0.4ex}
\bse
\be
\tr M(\lambda) = 2\cosh(S_\mathrm{i}(\lambda)/\epsilon)\,, 
\label{e:range1}
\ee
where $S_{\text{i}}(\lambda) = \int_{-L}^{L} \sqrt{-Z(x, \lambda)} \,\d x $.
Since $\tr M(\lambda)>2$ for all $\lambda$ in this range, these values of $\lambda$ are not part of the Lax spectrum. 

(ii) 
For $ \lambda \in (-q^{2}_\min, \infty)$, one has $Z(x,\lambda)>0$ for any $x\in[-L,L]$.  
Hence there are also no turning points in this case, and  
\vspace*{-0.4ex}
\be
\tr M(\lambda) = 2\cos(S_\mathrm{ii}(\lambda)/\epsilon)\,, 
\label{e:range2}
\ee
where $S_{\text{ii}}(\lambda) = \int_{-L}^{L} \sqrt{Z(x, \lambda)} \,\d x $.
Since $\tr M(\lambda)\leq 2 $ for all $\lambda$ in this range, these values of $\lambda$ form an infinitely long band.  

\begin{figure}[t!]
\centerline{\includegraphics[width=7.5cm]{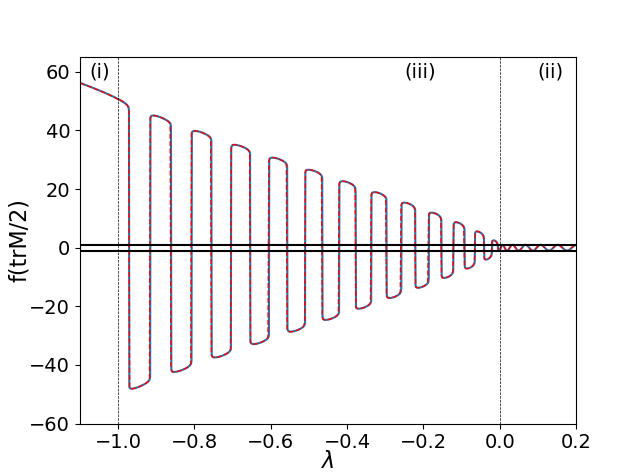}
\includegraphics[width=7.5cm]{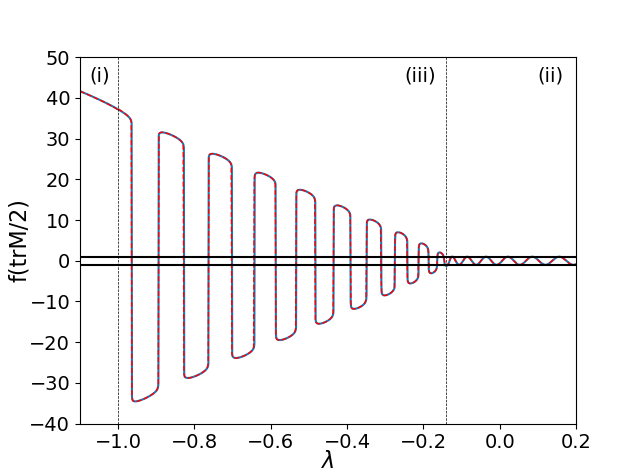}}
\centerline{\includegraphics[width=7.5cm]{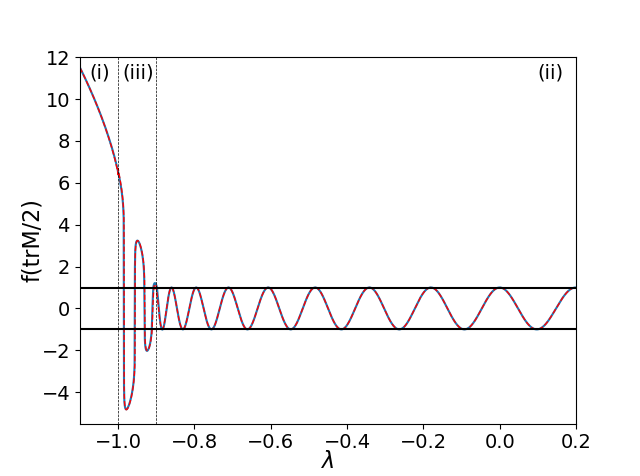}
\includegraphics[width=7.5cm]{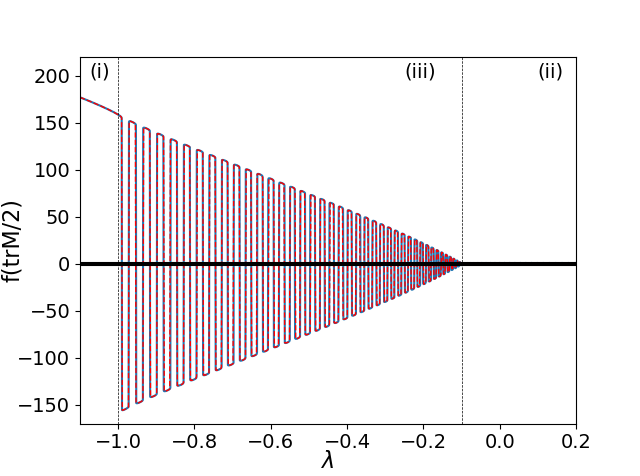}}
\caption{Trace of the monodromy matrix of the scattering problem.
\bb Top \eb left: the raised cosine potential given by Eq.~\eqref{e:IC1} with \bb$\epsilon=0.04$\eb. 
\bb Top right: the exp-sine potential given by Eq.~\eqref{e:IC2} with $\epsilon=0.0255$ \eb. 
\bb 
Bottom left: the dn potential given by Eq.~\eqref{e:IC3} with $m=0.1$ and $\epsilon = 0.05$.
Bottom right: the dn potential given by Eq.~\eqref{e:IC3} with $m=0.9$ and $\epsilon = 0.01$. 
\eb
Red (dashed): WKB approximation of $\tr M$ as a function of $\lambda$. 
Blue (solid): Results from numerical integration of the ODEs of the scattering problem Eq.~\eqref{e:LP1}. 
Dashed lines: the values $-\max[q^{2}(x)]$ and $-\min[q^{2}(x)]$, which define the boundaries of the three regions of the spectrum. 
Dot-dashed lines: the values $\tr M /2 = \pm 1$ which correspond to the edges of the spectrum.
Note that, since the amplitude of the oscillations grows exponentially, 
to capture the whole behavior in a single plot we take the vertical axis to be $f(\tr M/2)$ rather than $\tr M$ itself,
with the function $f(y)$ defined as $f(y) = y$ for $|y|\le1$ 
and $f(y) = \sgn(y)(1+\log_{10}|y|)$ for $|y|>1$, as in \cite{osbornebergamasco}.
}
\label{f:3}
\end{figure}

(iii) 
For $\lambda \in (-q^{2}_\max, -q^{2}_\min)$, there are two symmetric turning points, located at $x= \pm \tp(\lambda)$.
\bb
(That is, $\pm\tp(\lambda)$ are defined by the condition $Z(\pm \tp(\lambda), \lambda) = 0$.)
\eb
In this case one must write different representations for the eigenfunctions in each subregion and then 
connect the resulting expressions across the two transition regions.
The result of the analysis is (see section~\ref{s:wkbdetails} for further details)
\be
\tr M(\lambda) 
   = 2\cos(S_1(\lambda)/\epsilon)\cosh(2S_{2,\epsilon}(\lambda)/\epsilon)\,,
\label{e:tr}
\ee
\ese
where
\vspace*{-1ex}
\bse
\label{e:S1S2}
\begin{gather}
S_1(\lambda) = \int_{-\tp(\lambda)}^{\tp(\lambda)} \sqrt{|Z(x, \lambda)|} \,\d x\,,
\\
S_2(\lambda) = \int_{\tp(\lambda)}^{L} \sqrt{|Z(x, \lambda)|} \,\d x\,,
\end{gather}
\ese 
and $S_{2,\epsilon}(\lambda) = S_{2}(\lambda) + \epsilon \ln 2/2$.
Thus, in this region $\tr M$ is a rapidly oscillating function with exponentially growing amplitude as $\epsilon \downarrow  0$. 
Accordingly, this region is divided into a sequence of bands and gaps, and comprises the most interesting part of the Lax spectrum (see Fig.~\ref{f:3}).

In terms of the original spectral variable $\zeta$, the above results imply that the Lax spectrum is
comprised of the entire real axis plus the band $\zeta \in (-iq_\min, iq_\min)$,
together with bands and gaps for $ \zeta \in (-iq_\max, -iq_\min)\cup(iq_\min, iq_\max)$.
As shown in Fig.~\ref{f:2},
these predictions are in excellent agreement with the numerical results from the Floquet-Hill method 
for all of the potentials considered (see sections~\ref{s:numerics} and~\ref{s:wkbdetails} for further details).

Of course the WKB method yields not only bounds on the location of the spectrum, but also an asymptotic representation for the full monodromy matrix.
Therefore, one can further validate the WKB analysis by comparing its predictions with 
direct calculation of the monodromy matrix by numerical integration of the ordinary differential equations (ODEs) of the scattering problem,
Eq.~\eqref{e:LP1} (see section~\ref{s:numerics} for further details).
The results are shown in Fig.~\ref{f:3}, in which $\tr M$ is plotted as a function of $\lambda$ for 
the ICs in Eq.~\eqref{e:IC}.
(Equivalent results were obtained with other ICs, see section~\ref{s:numerics} for further details.)
As shown in the plots, the agreement is excellent in all three ranges of~$\lambda$.

\subsection{Effective solitons and soliton condensate}

Next we use the WKB expansion to identify the asymptotic properties of the spectral bands and gaps. 
Recall that the spectrum is composed of a sequence of bands and gaps, and that,
in the semiclassical limit, the gaps are confined to the region $\lambda \in (-q^{2}_{\max}, -q^{2}_{\min})$ [cf.\ Fig.~\ref{f:3}]. 
Again, here we limit ourselves to presenting the main results, referring the reader to section~\ref{s:wkbdetails}  
for some of the details.

\begin{figure}[t!]
\centerline{\includegraphics[width=7.5cm]{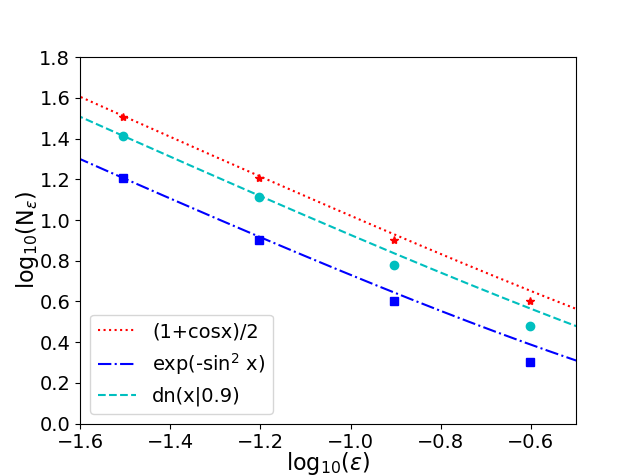}
\includegraphics[width=7.5cm]{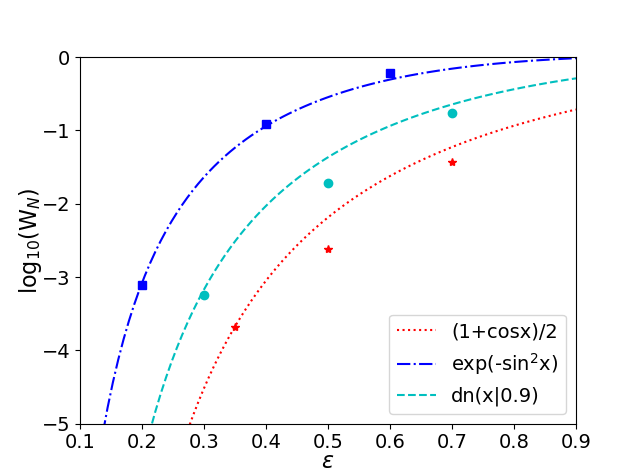}}
\caption{Left: The number of spectral bands 
as a function of $\epsilon$ for various single-lobe periodic potentials. The stars, circles and squares are data points obtained from direct computation of the monodromy matrix via numerical integration of the ZS scattering problem Eq.~\eqref{e:LP1}; the dotted, dashed and dot-dashed lines are the WKB predictions based on Eq.~\eqref{e:tr}. 
Right: The \bb $n_\epsilon$-th \eb relative bandwidth 
\bb(defined by Eq.~\eqref{e:bw})\eb\ as a function of $\epsilon$ for various single-lobe periodic potentials. The squares, circles and stars are data points obtained from direct numerical computation of the monodromy matrix via numerical integration of the ZS scattering problem Eq.~\eqref{e:LP1}; the dot-dashed, dashed and dotted curves are the WKB predictions based on \bb Eq.~\eqref{e:WN}.\eb 
}
\label{f:4}
\end{figure}

We first look at how the number of bands scales in the semiclassical limit. 
Let $N_{\epsilon}$ equal the number of spectral bands. 
Recall the WKB expansion of the trace function in the range $\lambda \in (-q^{2}_{\max},-q^{2}_{\min})$ in Eq.~\eqref{e:tr}. 
Because the amplitude of the oscillations grows exponentially, one has that, in this range, 
each spectral band is narrowly concentrated around one of the zeros of the trace. 
Hence the number of zeros $z_n$ of $\tr M$ is also the number of spectral bands. 
Using Eq.~\eqref{e:tr} and noting that $S_1(\lambda)$ is an increasing function, 
we see that $N_\epsilon$ is determined by the value of $S_1(\lambda)$ at
the edge of the infinitely long band, i.e, $\lambda = -q_\min^2$. 
That is,
to leading order, 
\bb the number of spectral bands is given by the expression \eb
\vspace*{-0.6ex}
\be
N_{\epsilon} = \Big\lfloor \frac{S_{1}(-q^2_\min)}{\pi \epsilon} + \frac{1}{2} \Big\rfloor\,,
\label{e:neps}
\ee
as $\epsilon\downarrow 0$, where 
\bm
the floor function $ \lfloor x \rfloor $ denotes the integer part of a real number $x$ (i.e., the largest integer less than or equal to $x$)\em.
This estimate for the number of bands can also be compared with the results obtained from direct numerical calculation of the monodromy matrix. 
The results, as shown in Fig.~\ref{f:4} (left), demonstrate that the asymptotic formula Eq.~\eqref{e:neps} matches the numerical results very well. 
Moreover, the asymptotic predictions become more accurate as $\epsilon \downarrow  0$ as expected.

Let $\lambda_n$ (for $n=1,2,\dots$) be the increasing sequence of values of $\lambda$ such that $\tr M = \pm 2$
(i.e., $\lambda_{4m-3}$ and $\lambda_{4m}$ are the values such that $\tr M =2$ and
$\lambda_{4m-2}$ and $\lambda_{4m-1}$ are the values such that $\tr M = -2$),
so the $n$-th spectral band is given by the interval $[\lambda_{2n-1},\lambda_{2n}]$.
The width of the $n$-th spectral band (which is approximately centered at $z_n$) 
and that of the $n$-th spectral gap are thus 
\vspace*{-0.4ex}
\be
w_{n} = \lambda_{2n} - \lambda_{2n-1},\qquad
g_{n} = \lambda_{2n+1} - \lambda_{2n},
\label{e:bandgapwidthdef}
\ee
respectively.
As in \cite{physd2016deng,pre2017deng}, one is also interested in the relative band width and relative gap width, as they can be used to distinguish solitonic excitations from nonsolitonic ones.
The relative band width and the relative gap width are defined respectively as%
\vspace*{-0.6ex}
\be
W_{n} = \frac{w_{n}}{w_{n}+g_{n}}\,,\qquad 
G_{n} = 1 - W_{n}.
\label{e:bw}
\ee
Using a Taylor expansion of Eq.~\eqref{e:tr} we get the following leading-order 
asymptotic expression of the $n$-th relative band width (see section~\ref{s:wkbdetails} for details):%
\vspace*{-0.6ex}
\be
W_{n} = \frac{4}{\pi} \sech \Big(\frac{2S_{2,\epsilon}(z_{n})}{\epsilon} \Big) 
\label{e:WN}
\ee
as $\epsilon \downarrow  0$. 
Again, one can compare these asymptotic expressions with the values obtained from direct numerical calculation of the monodromy matrix. 
The results, as shown in Fig.~\ref{f:4} (right), show excellent agreement between Eq.~\eqref{e:WN} and the numerical results.

The relative band width $W_n$ is a physically important quantity.  
This is because, as in the KdV and defocusing NLS equations
\cite{physd2016deng,prl2016deng,pre2017deng}, 
its value governs the characteristic features of periodic nonlinear excitations. 
More precisely, when $W_n \to 1$ the corresponding nonlinear excitation reduces to a constant background, 
whereas in the opposite limit, $W_n\to0$, the excitation becomes a soliton \bb(e.g., see chapter 5 in \cite{kamchatnov})\eb.

Accordingly, \bb given \eb a fixed threshold $\kappa\ll1$, we define a nonlinear excitation of the periodic problem to be an 
``effective soliton'' if its relative band width is less than~$\kappa$,
similarly to \cite{physd2016deng,prl2016deng,pre2017deng}.
Note that, while the introduction of an arbitrary threshhold parameter $\kappa$ might seem unsatisfactory,
we will show that the precise value of $\kappa$ is immaterial in the limit $\epsilon\downarrow0$.

The condition $W_n<\kappa$ provides a criterion that allows one to distinguish between
solitonic and non-solitonic excitations.
Explicitly, using the asymptotic expression in Eq.~\eqref{e:WN} for $W_n$, 
the inequality $W_n<\kappa$ implies that, as $\epsilon\downarrow0$,
the solitonic excitations are confined to the range $\lambda\in(-q^2_\max, \lambda_s)$,
where $\lambda_s$ is implicitly defined by the equation
\vspace*{-1ex}
\be
S_2(\lambda_s) = \frac{\epsilon}{2} \ln \Big( \frac{8}{\pi \kappa} \Big)\,.
\label{e:lambdasdef}
\ee
%
While no simple closed-form expression for $S_2(\lambda)$ or its inverse is available,
one can easily find $\lambda_s$ numerically. 
Also, one can obtain an analytical approximation for $\lambda_s$ by Taylor expanding $S_2(\lambda)$ near $\lambda=-q^2_\min$,
noting that $S_2(-q_\min^2)=0$.
Substituting the expansion into Eq.~\eqref{e:lambdasdef}, we obtain
\bb
that, to leading order, the spectral threshold of the solitonic excitations is given by 
\eb
\be
\lambda_{s,\mathrm{approx}} = \frac{\epsilon}{2S'_2(-q^2_\min)} \ln \Big( \frac{8}{\pi \kappa} \Big) - q^2_\min .
\label{e:lams_approx}
\ee
\bb
In other words, the band widths shrink exponentially with $\epsilon$ [as implied by~\eqref{e:WN}], but the gap widths and the solitonic threshold 
both scale linearly with $\epsilon$.
This is the same as what happens in the case of the KdV and defocusing NLS equations \cite{pre2017deng,physd2016deng,prl2016deng}.
\eb

The number $N_s$ of effective solitons equals the number of spectral bands of the trace function in the interval $(-q^2_\max, \lambda_s)$.
Using similar arguments as for Eq.~\eqref{e:neps}, we then immediately obtain
\vspace*{-0.4ex}
\be
N_s = \bigg\lfloor\frac{S_1(\lambda_s)}{\pi\epsilon} + \frac{1}{2} \bigg\rfloor\,.
\label{e:Nsoliton}
\ee
Moreeover, by expanding $S_1(\lambda)$ in a Taylor series about $\lambda=-q^2_\max$ 
[noting that $S_1(-q_\max^2)=0$] and substituting into Eq.~\eref{e:Nsoliton}, 
we can also obtain a linear approximation for $N_s$:
\be
N_{s,\mathrm{approx}} = \bigg\lfloor \frac{S_1'(-q_\max^2)}{\pi \epsilon}(q^2_\max - q^2_\min) + c \bigg\rfloor\,,
\ee
with 
$c = S'_1(-q^2_\max)/[2\pi S_2'(-q_\min^2)]\,\ln[8/(\pi\kappa)] + \frac12$.
Note that $N_{s,\mathrm{approx}}$ is independent of~$\kappa$ to leading order.
Hence the particular value chosen for the threshold~$\kappa$ becomes progressively less relevant as $\epsilon\bb \downarrow \eb 0$.

Importantly, note also from Eq.~\eqref{e:lams_approx} that $\lambda_s\to -q^2_\min$ as \bb $\epsilon\downarrow 0$ \eb.
This has an important practical consequence, 
since it means that all nonlinear excitations become effective solitons in the semiclassical limit.
Thus, \textit{the semiclassical limit of the focusing NLS equation with single-lobe periodic potential
is characterized by a coherent soliton condensate}.

\def\Ai{\mathop{\rm Ai}\nolimits}
\def\Bi{\mathop{\rm Bi}\nolimits}


\section{\uppercase{Numerical methods and further numerical results}}
\label{s:numerics}

In this section we provide some details about the numerical methods used and about the results presented in the previous sections.

\paragraph{Numerical solution of the focusing NLS equation.} 
All the numerical simulations of the semiclassical focusing NLS equation \eqref{e:NLS} were performed using an eighth-order Fourier split-step method~\cite{tappert,fornberg,yang,weideman} 
\bb
with at least $N = 2^{11}$ Fourier modes. 
The spatial accuracy of this method is spectral,
while the temporal accuracy is eighth-order.
The coefficients chosen for the time stepping are found by solving a system of algebraic equations (see \cite{yoshida} for details).
The time step was always chosen to satisfy the Courant-Friedrichs-Levy stability requirement \cite{yang,weideman}, namely,
$\Delta t \leq (\Delta x)^{2}/\epsilon$,
where $\Delta x = 2L/N$
and $2L$ is the spatial period of the particular IC considered.
(For the sech and Gaussian ICs discussed below, we took $2L = 30$.)
All results were checked for numerical convergence,
and the isospectral property of the scattering data was also checked 
using Floquet-Hill's method (see below), which served as further validation of 
numerical convergence. 
The corresponding simulations for each of the cases presented took several hours of computer time on a standard desktop computer.
All calculations were done in double precision.
\eb
\begin{figure}[t!]
\kern1ex
\centerline{\includegraphics[width=7.5cm]{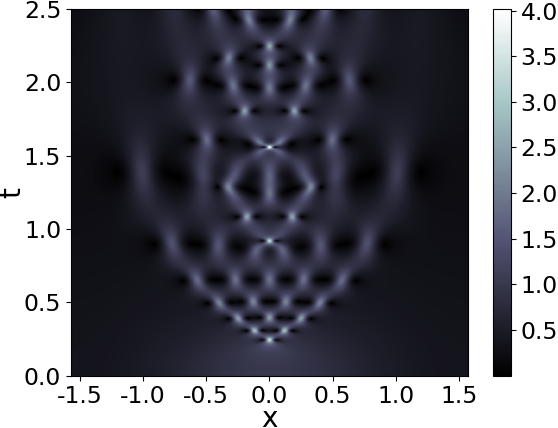}
\hspace{5mm} 
\includegraphics[width=7.5cm]{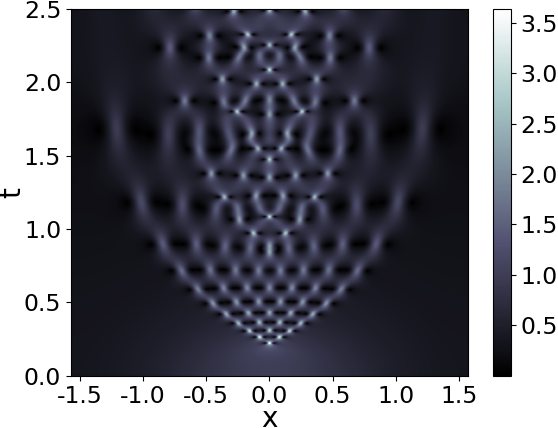}}
\vspace{2.5mm}
\centerline{\includegraphics[width=7.5cm]{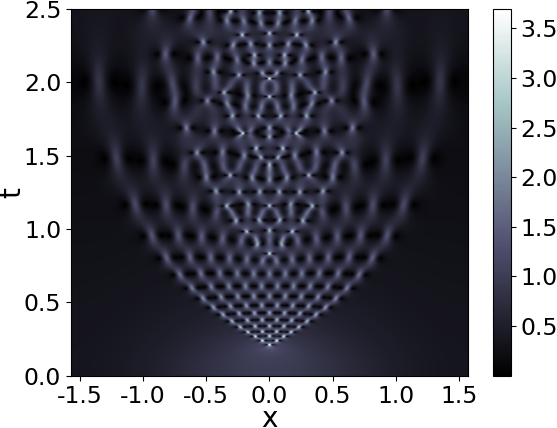}
\hspace{5mm}
\includegraphics[width=7.5cm]{numsol_expsine_eps0026_new_crop}}
\caption{\bb
Density plot of the amplitude $|q(x,t)|$ of the solution of the focusing NLS equation with the same ``exp-sine'' IC~\eqref{e:IC2} for decreasing values of the semiclassical parameter. 
Top left: $\epsilon = 0.078$. 
Top right: $\epsilon = 0.052$. 
Bottom left: $\epsilon = 0.037$. 
Bottom right: $\epsilon = 0.026$.
\eb}
\label{f:S1a}
\end{figure}

\paragraph{\bb Semiclassical dynamics and further numerical solutions. \eb} 
\bb
To illustrate the focusing dynamics of \eqref{e:NLS} as the semiclassical parameter tends to zero, Fig.~\ref{f:S1a} shows density plots of the amplitude $|q(x,t)|$ using the IC~\eqref{e:IC2} for decreasing values of $\epsilon$. Similar behavior was observed for the other potentials considered in this work (see Table~\ref{t1}). Note how the spatial period of the small-scale oscillations is proportional to $\epsilon$, but the location of the caustics becomes independent of $\epsilon$ as $\epsilon \downarrow 0$.
\eb

In Fig.~\ref{f:1} we showed for comparison purposes a solution with IC given by the following 
single-lobe potential on the infinite line:
\vspace*{-1ex}
\bse
\begin{gather}
q_\mathrm{sech}(x,0) = \sech x\,.
\label{e:IC5}
\end{gather}
Here we present additional numerical simulations of the focusing NLS equation~\eqref{e:NLS} with small dispersion and various kinds of ICs,
to investigate the generality of our results.
\bb
A list of ICs and the corresponding values of $\epsilon$ considered is given in Table~\ref{t1}.
\eb

\begin{table}[b!]
\begin{center}
\begin{tabular}{ |c|cccccc| }
\hline
$q(x,0)$  & \multicolumn6{c|}{$\epsilon$} 
\\
\hline
$(1+ \cos x)/2$ & 0.240 & 0.120 & 0.100 & 0.060 & 0.050 & 0.030 
\\
$\exp(-\sin^{2} x)$ & 0.100 & 0.078 & 0.052 & 0.037 & 0.0277 & 0.026 
\\
dn$(x|0.92)$  & 0.176 & 0.088 & 0.044 & 0.022 & 0.020 & 0.010 
\\
dn$(x|0.9)$   & 0.200 & 0.100 & 0.053 & 0.046 & 0.026 & 0.0255
\\
dn$(x|0.7)$   & 0.200 & 0.100 & 0.080 & 0.063 & 0.050 & 0.025 
\\
dn$(x|0.5)$   & 0.200 & 0.100 & 0.060 & 0.055 & 0.050 & 0.029 
\\
dn$(x|0.1)$   & 0.200 & 0.100 & 0.060 & 0.056 & 0.047 &  
\\
$\sech x$     & 0.200 & 0.100 & 0.050 & 0.042 & 0.037 & 0.020 
\\
$\exp(-x^{2})$ & 0.080 & 0.060 & 0.030 & 0.026 &  &  
\\
$1 - |x/\pi|$ & 0.160 & 0.080 & 0.055 & 0.040 & 0.030 & 0.019 
\\
$\exp(-|x|)$  &  0.120 & 0.060 & 0.030 & 0.027 & 0.014 &  
\\
\hline
\end{tabular}	
\end{center}
\caption{\bb List of ICs and the corresponding values of $\epsilon$ considered.\eb}
\label{t1}
\end{table}

\begin{figure}[t!]
\centerline{\includegraphics[width=7.5cm]{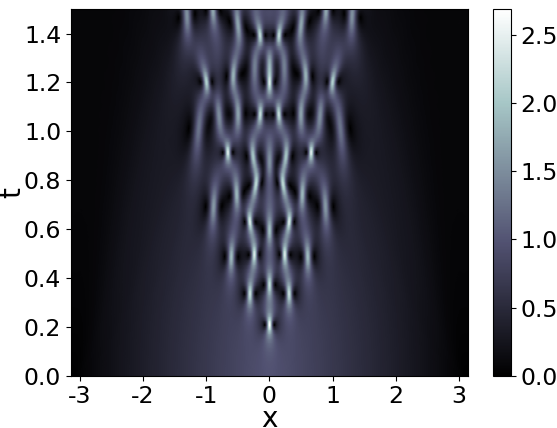}
\hspace{5mm} 
\includegraphics[width=7.5cm]{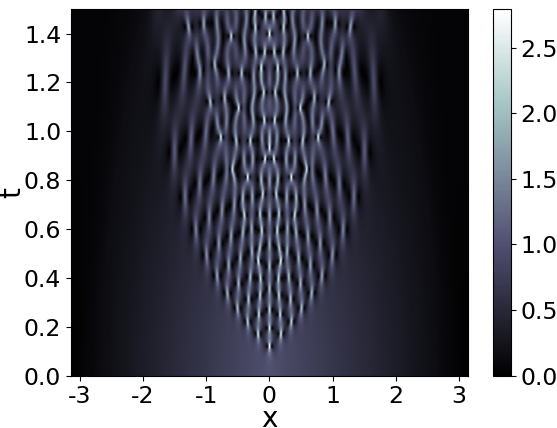}}
\vspace{2.5mm}
\centerline{\includegraphics[width=7.5cm]{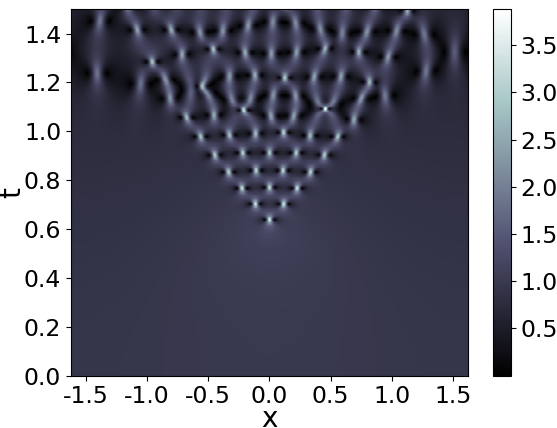}
\hspace{5mm}
\includegraphics[width=7.5cm]{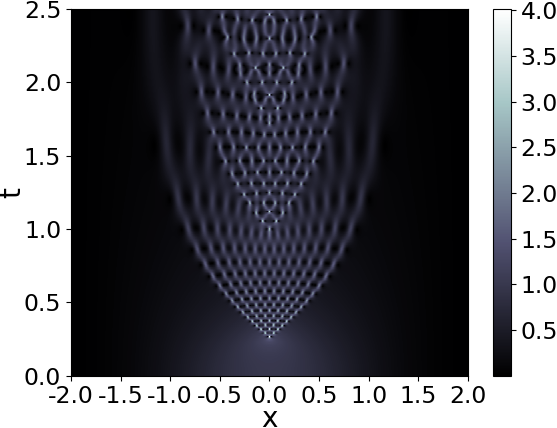}}
\caption{
\bb
Density plot of the amplitude $|q(x,t)|$ of the solution of the focusing NLS equation for various potentials and values of the semiclassical parameter. 
Top left: the ``tent-shape'' IC in Eq.~\eqref{e:ICpoint} with $\epsilon = 0.08$. 
Top right: the same ``tent-shape'' IC but with $\epsilon = 0.04$.
Bottom left: the ``dn'' IC in Eq.~\eqref{e:IC3} with $m=0.1$ and $\epsilon = 0.056$.
Bottom right: the Gaussian IC in Eq.~\eqref{e:IC4} with $\epsilon = 0.03$.
\eb}
\label{f:S1b}
\end{figure}

In Fig.~\ref{f:S1b} 
we present the results obtained from different kinds of ICs, in order to corroborate the general similarities between solutions with localized and periodic ICs. 
Specifically, we compare the solutions obtained with%
\vspace*{-1ex}
\begin{gather}
q_\mathrm{gaussian}(x,0) = \e^{-x^2}\,
\label{e:IC4}
\\
\noalign{\noindent 
as a potential on the infinite line,}
q_\mathrm{tent}(x,0) = 1 - |x/\pi|\,
\label{e:ICpoint}
\end{gather}
\ese
with $-\pi<x<\pi$,
as well as the dn IC in Eq.~\eqref{e:IC3} with other values of $m$, and $-K(m)<x<K(m)$.

Importantly, the results in Fig.~\ref{f:S1b} (top right) demonstrate that behavior similar to the one shown in Fig.~\ref{f:1} is produced
even by the non-differentiable IC~\eqref{e:ICpoint},
and virtually identical results were also obtained if the IC in Eq.~\eqref{e:ICpoint} is replaced by 
$q(x,0) = \e^{-|x|}$.
This is significant because the initial-value problem becomes elliptic in the limit $\epsilon\downarrow0$.
Therefore, analyticity of ICs is in general a necessary condition even just for solutions to exist,
and the problem becomes very sensitive to perturbations.
\bb
Prior numerical work by Bronski and Kutz \cite{bronskikutz1999} indicated an immediate detection by the dynamics of points of failure of analyticity of the data. 
This is confirmid by Fig.~\ref{f:S1b} (top right), which shows that the gradient catastrophe (i.e., the ``nose'' of the caustic) appears to 
develop almost immediately. 
\eb
On the other hand, Fig.~\ref{f:S1b} (top right) demonstrates that the resulting dynamical behavior is rather robust.
This is similar to what happens for the focusing NLS equation on the line with NZBC, where it was recently demonstrated that 
similar behavior occurs both with analytic and discontinuous data \cite{biondinimantzavinos,CPAM2017,biondinilimantzavinos}.

%
%
%

Some differences are evident in the behavior produced by Eq.~\eqref{e:IC3} with low values of $m$.
This should not be surprising, however, since \bb the function in \eb Eq.~\eqref{e:IC3} becomes shallower as $m$ decreases, and eventually tends to the constant value 1 $\forall x\in\Real$ as $m\to0$.
Therefore one should not expect the results to hold uniformly for all values of~$m$.
Nonetheless, the above numerical results provide further validation of the general nature of the behavior of solutions in the semiclassical limit with periodic or localized ICs.

\paragraph{Numerical calculation of the Lax spectrum via Floquet-Hill's method.} 

Recall that the ZS scattering problem is given by Eq.~\eqref{e:LP1}.
Since this problem is not self-adjoint, when calculating the spectrum numerically one must use 
techniques that are capable of efficiently calculating the spectrum in a large portion of the complex plane. 
One such technique is Floquet-Hill's method,
which applies Floquet-Bloch theory to give an almost uniform global approximation to the entire spectrum, 
as opposed to just an approximation of a few elements of the spectrum
(see \cite{DK2006} for details). 
Since $Q(x+2L) = Q(x)$, by Floquet's theorem all bounded solutions of Eq.~\eqref{e:LP1} are of the form 
\vspace*{-1ex}
\be
\phi(x, \zeta) = \e^{i\nu x}w(x, \zeta),
\label{e:flo} 
\ee
where $w(x+2L,\zeta) = w(x,\zeta)$, and $ \nu \in [0, \pi/L) $. 
As usual, we refer to $i\nu$ as the Floquet exponent. 
Inserting Eq.~\eqref{e:flo} into Eq.~\eqref{e:LP1} yields the modified eigenvalue problem
\be
\sigma_{3}[\epsilon(i \partial_{x} - \nu I) - iQ]w = \zeta w.
\label{e:fep}
\ee
While Eq.~\eqref{e:LP1} and Eq.~\eqref{e:fep} are obviously equivalent,
the crucial difference from a computational point of view is that, unlike $\phi(x,\zeta)$
the eigenfunction $w(x,\zeta)$ is also periodic.
One can therefore expand Eq.~\eqref{e:fep} in Fourier series to obtain
\be
\hat{\mf{L}}^{\epsilon}_\nu \hat{w} = \zeta \hat{w},
\label{e:operatoreigenvalue}
\ee
where $\hat{w} = (\ldots, \hat{w}_{-1}, \hat{w}_{0}, \hat{w}_{1}, \ldots)^{T}$ and $\hat{w}_{j}$ is the $j$-th Fourier coefficient of $w(x,\zeta)$, and 
$$
\hat{\mf{L}}^{\epsilon}_\nu = \begin{pmatrix} -\epsilon(k + \nu\,I)  & -i \mf{T} \\ -i \mf{T} & \epsilon(k + \nu\,I) \end{pmatrix},
$$
$k = \diag(k_n)_{n\in\Integer}$ is the doubly infinite diagonal matrix of Fourier wavenumbers, with $k_n = n \pi/L$, 
and $\mf{T}$ is the doubly infinite Toeplitz matrix representing the convolution operator that is produced by the Fourier series of $q(x)w(x,\zeta)$.

The method then approximates the eigenvalues of the scattering problem
by numerically computing the eigenvalues of the finite matrix obtained by a truncation of Eq.~\eqref{e:operatoreigenvalue}.
The numerical accuracy of the approximation is dependent on the number of Fourier modes used and on the eigenvalue solver. 
Note also that the density of the spectral bands depends on number of Floquet exponents chosen in the interval $[0, \pi/L)$.  
\bb 
For each of the Floquet-Hill's method simulations shown in this work we used no less than $N=2^{8}$ Fourier modes and at least $10^{4}$ Floquet exponents. 
\eb
All results were checked for numerical convergence.
\bb
Namely, we ensured that the number of Fourier modes and the step size for the Floquet exponent were such that the results were independent of the specific values of each.  
We also double-checked our results with exactly solvable examples such as the step, plane wave and sech potentials.  
\eb

\begin{figure}[t!]
	\centerline{\includegraphics[width=7.5cm]{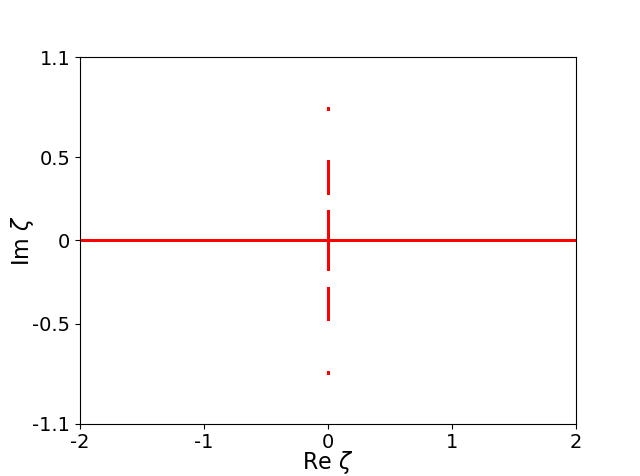}
		\includegraphics[width=7.5cm]{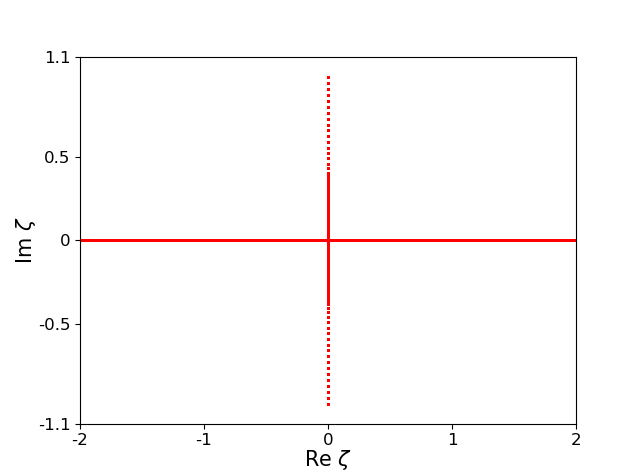}}
	\centerline{\includegraphics[width=7.5cm]{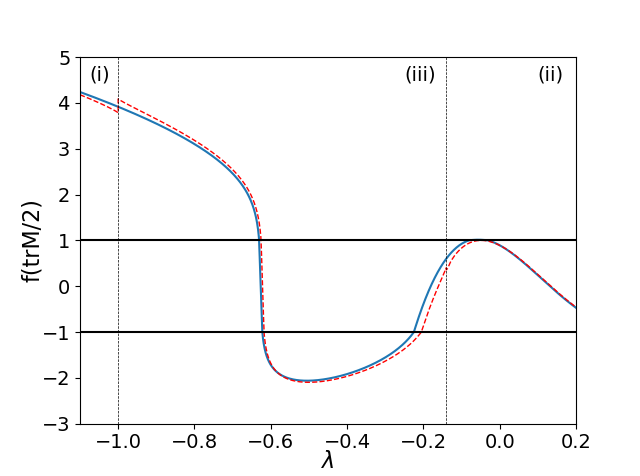}
		\includegraphics[width=7.5cm]{wkb_expsine_eps00255_neww}}
	\caption{Top row: Spectrum (red bands) of the ZS scattering problem as computed numerically via Floquet-Hill's method. 
	Left column: exp-sine potential in Eq.~\eqref{e:IC2}, with $\epsilon = 0.3$. 
	Right column: same potential with $\epsilon = 0.0255$. 
	Bottom row: trace of the monodromy matrix. 
	Red dashed curves: WKB approximation of $\tr M$ as a function of $\lambda$. 
	Blue solid curves: Results from numerical integration of the scattering problem. 
	Dashed lines: the values -max[$q^{2}(x)$] and -min[$q^{2}(x)$] that define the boundaries of the three regions of the spectrum. Solid black lines: the values $\tr M/2 = \pm 1$ corresponding to the spectral band edges.
	}
	\label{f:S2}
\end{figure}

Additional plots of the numerical calculation of the Lax spectrum are provided in Fig.~\ref{f:S2}. 
The top row shows numerical calculations of the Lax spectrum via Floquet-Hill's method. 
Note how, as $\epsilon \downarrow 0$ the spectral data clusters on the real and imaginary axes, the number of bands grows, and the band widths decay to resemble point spectra. 
The bottom row shows the WKB approximation of $\tr M(\lambda)$, where $\lambda = \zeta^{2}$. We see excellent agreement between the WKB approximation and Floquet-Hill's method, especially as $\epsilon \downarrow 0$, as expected.

\paragraph{Numerical calculation of the monodromy matrix.}
The results obtained from the Floquet-Hill method described above, and the predictions obtained from the WKB expansion of the scattering problem (see below), 
can both be tested by comparing them with the results of direct numerical integration of the scattering problem.

Recall that the monodromy matrix is defined by Eq.~\eqref{e:Mdef} as
$M(\zeta) = \Phi(x-L,\zeta)^{-1}\Phi(x+L,\zeta)$,
where $\Phi(x,\zeta)$ is any fundamental matrix solution of Eq.~\eqref{e:LP1}. 
Choosing $\Phi(0,\zeta) = I$, \bm where $I$ is the $2\times 2$ identity matrix \em as IC, 
one can obtain the monodromy matrix simply as
\vspace*{-1ex}
\be
M(\zeta) = \Phi(2L,\zeta).
\label{e:Msimpledef}
\ee
Integrating Eq.~\eqref{e:LP1} numerically using a fourth-order Runge-Kutta method with step size $\Delta x \leq 10^{-3}$ 
then allows one to compute the monodromy matrix via Eq.~\eqref{e:Msimpledef}. 
Since $\tr M$ yields all the necessary information about the spectrum of the scattering problem, 
one can therefore use it to validate the result that the spectral bands converge to the real and imaginary $\zeta$-axes in the semiclassical limit
as well as the asymototic expressions for the location of the spectral bands 
(thus confirming the results obtained with the WKB method).

\section{\uppercase{WKB expansions and asymptotic calculations}}
\label{s:wkbdetails}

In this section we provide some details of the asymptotic calculation of the trace of the monodromy matrix via the WKB method. 

\paragraph{Eikonal and transport equations.}

Recall that the change of variables $v = \phi_{1} + i\phi_{2}$, and $\bar{v} = \phi_{1} - i\phi_{2}$ transforms the scattering problem Eq.~\eqref{e:LP1} 
into the time-independent Schr\"odinger equation Eq.~\eqref{e:ode}.

We look for an asymptotic representation of solutions of the second-order differential equation~\eqref{e:ode} in the form
\vspace*{-0.4ex}
\be
v(x) = (A(x) + O(\epsilon))\e^{iS(x)/\epsilon},\qquad \epsilon \bb \downarrow \eb 0\,.
\label{e:ansatz}
\ee
Substituting Eq.~\eqref{e:ansatz} into Eq.~\eqref{e:ode} yields
the eikonal and transport equation, respectively, as
\vspace*{-1ex}
\bse
\label{e:eikonaltransport}
\begin{gather}
(S')^{2} = Z(x, \lambda)\,,
\\
2S'(x)A' + S''(x)A + q'(x)A = 0\,.
\end{gather}
\ese
\bb
These equations can be easily integrated (up to arbitrary additive and multiplicative constants, respectively)
once the sign of $Z(x,\lambda)$ is known.
\eb
Because of the possible presence of turning points however, we need to analyze the spectrum in three separate ranges of values of~$\lambda$.

\begin{figure}[t!]
\centerline{\includegraphics[width=7.5cm]{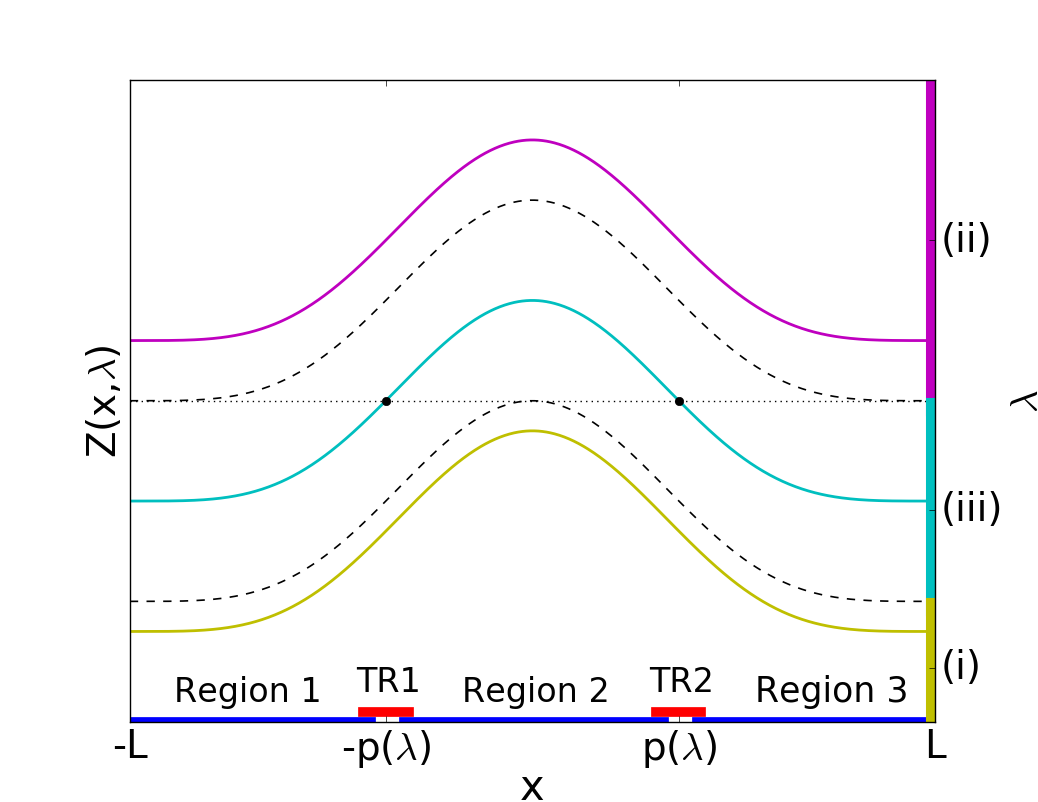}}
\caption{Schematic plot of $Z(x, \lambda) = \lambda + q^{2}(x)$ for a single-lobe periodic potential $q(x)$ with $ \lambda \in (-q^{2}_{\min}, \infty) $ (purple, range (ii)), $\lambda=-q^{2}_{\min}$ (black dashed), $ \lambda \in (-q^{2}_{\max}, -q^{2}_{\min}) $ (light blue, range (iii)), $\lambda = -q^{2}_{\max}$ (black dashed), $ \lambda \in (-\infty, -q^{2}_{\max}) $ (yellow, range (i)), and $ Z(x, \cdot) = 0 $ (black dotted). For the WKB analysis we have regions 1-3 (blue lines) and transition regions 1-2 (red lines). The overlap in these regions allow for asymptotic matching. For $\lambda$ in range (iii) we have $x = \pm \tp(\lambda)$ are the turning points.}
	\label{f:S4}
\end{figure}

\paragraph{Range (i): $\lambda < -q^{2}_{\max}$.}
For $\lambda$ in this range, one has $Z(x, \lambda) < 0$ (cf.\ Fig.~\ref{f:S4}), and the \bb leading order WKB approximations \eb are of the form
\bse
\begin{gather}
v_{\bb\pm\eb}(x,\lambda) = A_{\pm}(x)\e^{\bb S_{\bb \mp \eb} (x)\eb / \epsilon},
\label{e:vpm1def}
\\
\noalign{\noindent\bb with\eb}
\bb 
S_{\pm}(x) = \pm \int_{-L}^x \sqrt{|Z(x, \lambda)|} \,\d x\,,
\eb
\label{e:Spm1def}
\\
A_{\pm}(x) = \frac{\sqrt{\mp i \sqrt{|Z(x, \lambda)|} + q(x)}}{\sqrt[4]{|Z(x, \lambda)|}}.
\label{e:Apm1def}
\end{gather}
\ese
Thus, a fundamental matrix solution in range~(i) is given by
\be
\Phi(x, \lambda) = \begin{pmatrix} v_{-} & v_{+} \\ v'_{-} & v'_{+} \end{pmatrix}.
\label{e:FMS}
\ee
Since $Z(x, \lambda) \neq 0$ in this range, this solution is valid for all $ x \in [-L,L]$. 
We can obtain the monodromy matrix from Eq.~\eqref{e:Mdef} evaluated at $x=-L$.
Simple matrix algebra then gives the trace of $M$ as Eq.~\eqref{e:range1}. 


\paragraph{Range (ii): $\lambda > -q^{2}_{\min}$.}
For $\lambda$ in this range, one has $Z(x, \lambda) > 0$ (cf.\ Fig.~\ref{f:S4}), and the \bb leading order WKB approximations \eb are of the form 
\bse
\label{e:vpmdef2}
\begin{gather}
v_{\pm}(x) = A_{\pm}(x)\e^{iS_{\pm}(x) / \epsilon}, 
\\
\noalign{\noindent \bb with\eb}
\bb
S_{\pm}(x) = \pm \int_{-L}^x \sqrt{Z(x, \lambda)} \,\d x\,,
\eb
\label{e:S}
\\
\bb
A_{\pm}(x) = \frac{\sqrt{\mp \sqrt{|Z(x, \lambda)|} + q(x)}}{\sqrt[4]{|Z(x, \lambda)|}},
\eb
\label{e:A}
\end{gather}
\ese
Thus, we again have that a fundamental matrix solution in range~(ii) is given by Eq.~\eqref{e:FMS}, but with 
$v_\pm(x,\lambda)$ now given by Eq.~\eqref{e:vpmdef2}.
Since $Z(x, \lambda) \neq 0$ in this range as well, the above solution is also valid for all $x \in [-L, L]$. 
Thus, as before, we obtain the monodromy matrix from Eq.~\eqref{e:Mdef} at $x=-L$.
Simple matrix algebra then yields the trace of $M$ as Eq.~\eqref{e:range2}.

\paragraph{Range (iii): $-q^{2}_{\max} < \lambda < -q^{2}_{\min}$.}
For $\lambda$ in this range, $Z(x, \lambda)$ has two real zeros at $x = \pm \tp(\lambda)$, i.e.,
\be
Z(\pm \tp(\lambda), \lambda) = 0\,,
\ee
(cf.\ Fig.~\ref{f:S4}).  
Thus, in the context of WKB there are two real turning points, one at each zero of $Z(x, \cdot)$. 
We must therefore discuss the behavior of the WKB \bb approximation \eb in the following five subregions of the fundamental period $ x \in [-L,L] $:
\begin{itemize}
\advance\itemsep-5pt
\item[(a)] Region 1, $ x \in [-L, -\tp(\lambda)) $.
\item[(b)] Transition 1, $ x \in (-\tp(\lambda)-\delta, -\tp(\lambda)+\delta)$, $\; \delta > 0 $.
\item[(c)] Region 2, $ x \in (-\tp(\lambda), \tp(\lambda)) $.
\item[(d)] Transition 2, $ x \in (\tp(\lambda)-\delta, \tp(\lambda)+\delta)$, $\; \delta > 0$.
\item[(e)] Region 3, $ x \in (\tp(\lambda), L] $.
\end{itemize} 
These regions are shown in Fig.~\ref{f:S4}.
For brevity we drop the $\lambda$ dependence of the turning points and simply write $\tp = \tp(\lambda)$.
Note that one could exploit the evenness and reality of the potential, and the resulting symmetries of the eigenfunctions, 
to obtain the eigenfunctions for $x<0$ in terms of those for $x>0$.  
Namely, $\Phi(x,\zeta) = \sigma_1\Phi(-x,\zeta)\sigma_1$, where $\sigma_1$ is the first Pauli matrix.

\paragraph{Region 1.}
The WKB approximation for the general solution of Eq.~\eqref{e:ode} in this region is 
\vspace*{-0.6ex}
\be
v_{1}(x) = a_{1}^{+}v_{+}(x) + a_{1}^{-}v_{-}(x),
\label{e:27}
\ee
where 
$v_\pm(x)$ are given by Eq.~\eqref{e:vpm1def},
\bb
with the lower integration limit replaced by $-p$
in Eq.~\eqref{e:Spm1def}
(to avoid any issues related to the sign change of $Z(x,\lambda)$)
\eb
and $A_\pm(x)$ given by Eq.~\eqref{e:Apm1def}.

\paragraph{Transition region 1.} 
The first transition region corresponds to a neighborhood of the first transition point, $x= -\tp$.
In this region we have that $Z(x, \lambda) = a(x+\tp) + o(1)$ as $ x \to -\tp $, with $ a > 0 $. 
Following \bb the \eb standard approach [e.g., see \cite{bender}], one can then obtain the solution of Eq.~\eqref{e:ode} in this region to leading order as
\be
v_{1 \to 2}(x) = c_{1}^{-} \Ai[\xi(x, \lambda)] + c_{1}^{+} \Bi[\xi(x, \lambda)],
\ee
where $\xi(x, \lambda) = -a^{1/3}(x+\tp)/\epsilon^{2/3}$ and $\Ai(\cdot)$ and $\Bi(\cdot)$ are the Airy functions \cite{NIST}.

\paragraph{Region 2.}
The WKB \bb approximation to the solution \eb of Eq.~\eqref{e:ode} in this region has two different but equivalent representations depending on the starting point of integration, namely:
\bse
\begin{gather}
v_{2}(x) = a_{2}^{+}v_{+}(x) + a_{2}^{-}v_{-}(x)\,,
\\
 \hspace{-0.5mm} \bar{v}_{2}(x) = \bar{a}_{2}^{+}\bar{v}_{+}(x) + \bar{a}_{2}^{-}\bar{v}_{-}(x)\,,
\end{gather}
\ese
where
\bse
\begin{gather}
\hspace{-1.4mm} v_{\pm}(x) = A_{\pm}(x) \exp \Big(\pm i \int_{-\tp}^{x} \sqrt{|Z(s, \lambda)|} \d s / \epsilon \Big)\,,
\\
\hspace{-4.2mm} \bar{v}_{\pm}(x) = A_{\pm}(x) \exp \Big( \pm i \int_{\tp}^{x} \sqrt{|Z(s, \lambda)|} \d s / \epsilon \Big)\,,
\end{gather}
\ese
and 
$A_\pm(x)$ given by Eq.~\eqref{e:A}.

\paragraph{Transition region 2.} 
In the second transition region 
we have $ Z(x, \lambda) = -b(x-\tp) + o(1)$ as $ x \to \tp $, with $b>0$. 
Following similar steps as before, one can write the solution of Eq.~\eqref{e:ode} in this region to leading order as
\be
v_{2 \to 3}(x) = c_{2}^{-} \Ai[\eta(x, \lambda)] + c_{2}^{+} \Bi[\eta(x, \lambda)],
\ee
where $\eta(x, \lambda) = b^{1/3}(x - \tp)/\epsilon^{2/3} $. 

\paragraph{Region 3.} 
The WKB solution of Eq.~\eqref{e:ode} in this region is
\be
v_{3}(x) = a_{3}^{+}(x)v_{+}(x) + a_{3}^{-}v_{-}(x),
\ee
where 
\bb
$v_\pm(x)$ are as in Eq.~\eqref{e:vpm1def} 
and the lower integration limit in Eq.~\eqref{e:Spm1def} replaced by $p$.
\eb

\paragraph{Asymptotic matching and connection formulae.} 
We now perform asymptotic matching across each boundary layer. 
We begin by matching $v_{1}(x)$ with $v_{1\to2}(x)$. 
To leading order, in region~1 one has
%
\vspace*{-0.6ex}
\be
v_\pm(x) = 
 \frac{\sqrt[4]{|\lambda|}}{\sqrt[4]{a|x+\tp|}} \e^{\pm\frac{2}{3}a^{1/2}|x+\tp|^{3/2}/\epsilon}\,,
\quad x\to -\tp^-\,.
\ee
Using the well-known asymptotic expansions of the Airy functions \bb[cf.\ section 9.7 in \cite{NIST}] \eb and 
requiring that the expansion for $v_1(x)$ as $x\to-\tp^-$ matches that of $v_{1\to2}(x)$ as $\xi\to\infty$
we obtain the connection formula
\vspace*{-0.6ex}
\be
\begin{pmatrix} c_{1}^{-} \\ c_{1}^{+} \end{pmatrix} = C_{1} \begin{pmatrix} a_{1}^{-} \\ a_{1}^{+} \end{pmatrix}, \;\; C_{1} = \frac{\sqrt[4]{\pi^{2}|\lambda|}}{(a\epsilon)^{1/6}} \begin{pmatrix} 2 & 0 \\ 0 & 1 \end{pmatrix}.
\ee
Next, we match $v_{1\to2}(x) $ with $v_{2}(x)$. 
To leading order, 
in region~2 one has
\vspace*{-1ex}
\be
v_\pm(x) = 
\frac{\sqrt[4]|\lambda|}{\sqrt[4]{a(x+\tp)}} \e^{\pm i\frac{2}{3}a^{1/2}(x+\tp)^{3/2}/\epsilon},
\qquad
x\to - \tp^+\,.
\ee
Requiring that the above expansion for $v_2(x)$ matches that for $v_{1\to2}(x)$ as $\xi\to-\infty$ 
we obtain the connection formula
\be
\begin{pmatrix} a_{2}^{+} \\ a_{2}^{-} \end{pmatrix} = C_{2} \begin{pmatrix} c_{1}^{-} \\ c_{1}^{+} \end{pmatrix}, \; \; C_{2} = \frac{(a\epsilon)^{1/6}}{2\sqrt[4]{\pi^{2}|\lambda|}} \begin{pmatrix} -i\e^{i\pi/4} & \e^{i\pi/4} \\ i\e^{-i\pi/4} & \e^{-i\pi/4} \end{pmatrix}.
\ee
Similarly,
matching $v_{2}(x)$ with $\bar{v}_{2}(x)$ yields the
connection formula%
\be
\begin{pmatrix} \bar{a}_{2}^{+} \\ \bar{a}_{2}^{-} \end{pmatrix} = C_{3} \begin{pmatrix} {a}_{2}^{+} \\ {a}_{2}^{-} \end{pmatrix}, \; \; C_{3} = \e^{i \sigma_{3} \int_{-\tp}^{\tp} \sqrt{|Z(s, \lambda)|} \d s/\epsilon},
\ee
where $\sigma_{3} = \text{diag}(1,-1)$. 
Next, 
matching $\bar{v}_{2}(x)$ with $v_{2\to 3}(x)$,
we obtain
\be
\begin{pmatrix} c_{2}^{+} \\ c_{2}^{-} \end{pmatrix} = C_{4} \begin{pmatrix} \bar{a}_{2}^{+} \\ \bar{a}_{2}^{-} \end{pmatrix}, \;\; C_{4} = \frac{\sqrt[4]{\pi^{2}|\lambda|}}{(b\epsilon)^{1/6}} \begin{pmatrix} \e^{i\pi/4} & \e^{-i\pi/4} \\ -i\e^{i\pi/4} & i\e^{-i\pi/4} \end{pmatrix}.
\ee
Finally, 
matching $v_{2\to 3}(x)$ with $v_{3}(x)$
we get
\be
\begin{pmatrix} a_{3}^{-} \\ a_{3}^{+} \end{pmatrix} = C_{5} \begin{pmatrix} {c}_{2}^{+} \\ {c}_{2}^{-} \end{pmatrix}, \;\; C_{5} = \frac{(b\epsilon)^{1/6}}{\sqrt[4]{\pi^{2}|\lambda|}} \begin{pmatrix} 1 & 0 \\ 0 & 1/2 \end{pmatrix}.
\ee
Combining all of the above expressions we obtain that the matrix%
\be
C = C_{5}C_{4}C_{3}C_{2}C_{1}
\label{e:cmatrix}
\ee
allows us to extend a solution in region~1 to one in region~3. 

\bb
Some remarks are now in order.  It is well known that, in general, one must deal with the directional character of the WKB method when 
connecting through \bm classically \em forbidden regions \cite{bender,Berry,miller2006}.  
Note that our calculations to obtain the connection formulae are purely formal.  
Moreover, the approach we employed is the time-honored method of matching asymptotic expansions.
Indeed, the approach we used is exactly the same as the one used in \cite{ablowitzfokas,holmes2013,simmondsmann,hinch}, where connection problems of exactly the same kind were presented and solved in exactly the same way.
In any case, the asymptotic expression we obtained for the trace of the monodromy matrix agrees extremely well with the results of direct numerical simulations of the spectrum of the scattering problem (cf. Figs.~\ref{f:3} and \ref{f:S2}), and also agrees very well with the results of Floquet-Hill’s method.  All of this serves as a strong validation of the WKB results.
Finally, exactly the same approach was already applied with similar success to characterize the spectrum of the time-independent Schrodinger equation for the defocusing Zakharov-Shabat scattering problem in \cite{pre2017deng,physd2016deng}.
\eb

\begin{figure}[t!]
\centerline{\includegraphics[width=7.5cm]{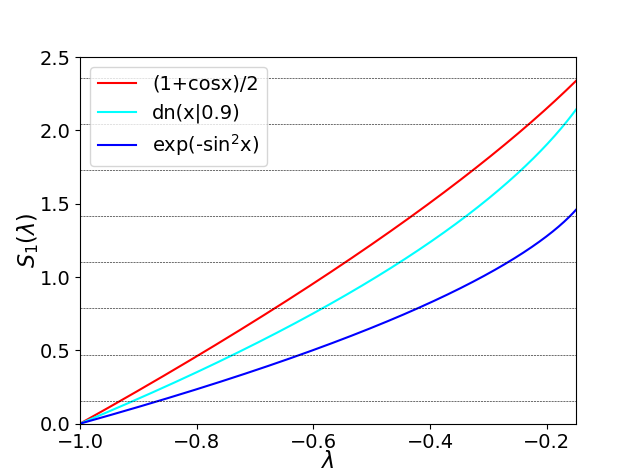}
	\includegraphics[width=7.5cm]{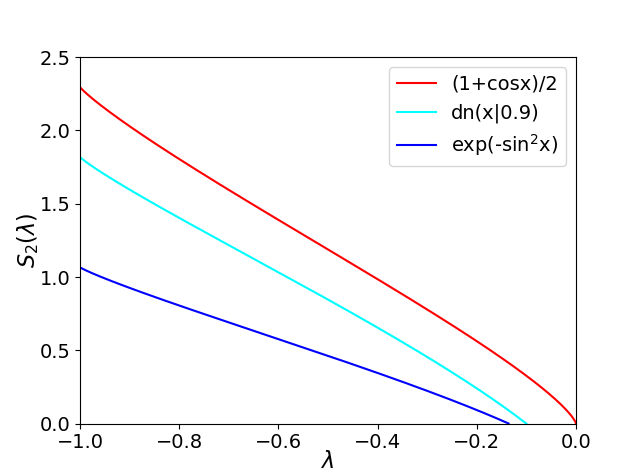}}
	\caption{Left: $S_{1}(\lambda)$ for Eq.~\eqref{e:IC1} (red), Eq.~\eqref{e:IC2} (blue), Eq.~\eqref{e:IC3} with $m=0.9$ (light blue), and $(n - 1/2)\pi \epsilon$ (horizontal dashed). Right: $S_{2}(\lambda)$ for Eq.~\eqref{e:IC1} (red), Eq.~\eqref{e:IC2} (blue), and Eq.~\eqref{e:IC3} with $m=0.9$ (light blue).
	} 
	\label{f:S5}
\end{figure}

\paragraph{Monodromy matrix in range~(iii).}
We now have all the necessary information to calculate the trace of $M$ in range (iii). 
To simplify the resulting expressions,  it is convenient to introduce the function
\be
S(x, \lambda) = \int_{-\tp}^{x} \sqrt{|Z(s, \lambda)|} \d s\,,
\ee
as well as $S_1(\lambda)$ and $S_2(\lambda)$ defined in Eq.~\eqref{e:S1S2}.
Note that $S_{1}$ is a nonnegative monotone increasing function of $\lambda$ in $(-q^{2}_\max, -q^{2}_\min)$, 
while $S_{2}$ is a nonnegative monotone decreasing function of $\lambda$ in the same domain.
A plot of both functions is shown in Fig.~\ref{f:S5}.

We can write a fundamental matrix solution of the scattering problem in the form of Eq.~\eqref{e:FMS}, where
$v_{\pm}(x)$ are given by Eq.~\eqref{e:vpm1def} in region~1, 
and by their continuation (obtained through the connection formulae discussed above) for $x\in(-\tp(\lambda),L]$.
Explictly, to leading order we have
\vspace*{-0.6ex}
\bse
\begin{gather}
\Phi(-L, \lambda) = \Phi_o\,\e^{-S_2(\lambda)\sigma_3/\epsilon}\,,
\\
\Phi(L, \lambda) =  \Phi_o\,\e^{S_2(\lambda)\sigma_3/\epsilon}\,C\,,
\end{gather}
\ese
where 
\be
\Phi_o = 
  \begin{pmatrix} A_-(L) & A_+(L) \\
     A_-(L)\sqrt{|Z(L, \lambda)|}/\epsilon & -A_+(L)\sqrt{|Z(L, \lambda)|}/\epsilon 
  \end{pmatrix}
\ee
and 
$C$ is the overall connection matrix given by Eq.~\eqref{e:cmatrix}. 
The monodromy matrix can then again be computed via Eq.~\eqref{e:Mdef}.
Simple matrix algebra then gives that the trace of $M$ is given by Eq.~\eqref{e:tr}.

\paragraph{Number of bands, band widths and gap widths.}
We now provide some details of the calculations used to find the asymptotic expressions for the band widths, gap widths and number of bands. 
We begin by finding an asymptotic representation for the number of bands. 
From Eq.~\eqref{e:tr} it is clear that the oscillation amplitude grows exponentially as $\epsilon \downarrow 0$. 
This means that the values of $\lambda$ such that $\tr M = \pm 2$ (which are the band and gap edges) are clustered near the zeros $z_{n}$ of $\tr M$. 
In turn, the zeros are given by the equation
\vspace*{-0.4ex}
\be
S_{1}(z_{n}) = (n - 1/2)\pi \epsilon, \hspace{3mm} z_{n} \in (-q^{2}_{\max}, -q^{2}_{\min})\,.
\label{e:zz}
\ee
Then, since $S_{1}(\lambda) $ is a monotonically increasing function (see Fig.~\ref{f:S5}), one obtains Eq.~\eqref{e:neps}.
\bb
Note that Eq.~\eqref{e:zz} is equivalent to the Bohr-Sommerfeld quantization condition that one would obtain for the discrete eigenvalues of a potential well 
by taking into account the directional character of the WKB approximation (e.g., see \cite{bender,miller2006}).
\eb

Next recall that the $n$-th relative band width is defined by Eq.~\eqref{e:bw} as
$W_{n} = {w_{n}}/({w_{n}+g_{n}})$,
where the absolute band width and gap width are given by Eq.~\eqref{e:bandgapwidthdef},
and $\lambda_n$ denotes the increasing sequence of values of $\lambda$ such that $\tr M = \pm 2$. 
It is convenient to introduce the half-trace as $\tau(\lambda) = \tr M(\lambda)/2$. 
Taylor expanding $\tau$ about $z_{n}$ and differentiating, we have
\vspace*{-1ex}
\be
\tau(\lambda) = \tau'(\lambda-z_{n}) + \frac{\tau''}{2}(\lambda-z_{n})^{2} + O(\lambda-z_{n})^{3},
\label{e:taylor}
\ee
as $\lambda \to z_{n}$ and
\vspace*{-1ex}
\begin{align*}
\tau' \big|_{\lambda = z_{n}} &= -\frac{S_{1}'(z_{n})}{\epsilon}\cosh(2S_{2,\epsilon}(z_{n})/\epsilon)(1+o(1)),  \\ 
\tau'' \big|_{\lambda = z_{n}} &= \frac{1}{\epsilon^{2}} \e^{2S_{2}(z_{n})/\epsilon}(1+o(1)),
\end{align*}
as $\epsilon \downarrow 0$. Evaluating  Eq.~\eqref{e:taylor} at $\lambda_{2n-1}$ yields 
\vspace*{-1ex}
\be
\lambda_{2n-1} - z_{n} = 1/\tau' + O(\epsilon \e^{-4S_{2}(z_{n})/\epsilon}), \hspace{2mm} \epsilon \downarrow 0.
\ee
Thus,
\vspace*{-1ex}
\be
w_{n} = \frac{2\epsilon}{|S_{1}'(z_{n})|} \text{sech} \Big(\frac{2S_{2,\epsilon}(z_{n})}{\epsilon} \Big) + O(\epsilon \e^{-4S_{2}(z_{n})/\epsilon}),
\label{e:wn}
\ee
as $\epsilon \downarrow 0 $. 
Next, note that since
\vspace*{-1ex}
\begin{align*}
w_{n} + g_{n} &= (\lambda_{2n} - z_{n}) + (z_{n} - \lambda_{2n-1}) + (\lambda_{2n+1} - \lambda_{2n}), \\ z_{n+1} - z_{n} &= (\lambda_{2n} - z_{n}) + (\lambda_{2n+1} - \lambda_{2n}) + (z_{n+1} - \lambda_{2n+1}),
\end{align*}
we have 
\vspace*{-1ex}
\be
(w_{n} + g_{n}) - (z_{n+1}-z_{n}) = O(\epsilon \e^{-4S_{2}(z_{n+1})/\epsilon}), \hspace{2mm} \epsilon \downarrow 0.
\nonumber
\ee
From Eq.~\eqref{e:zz} we also have $S_{1}(z_{n+1}) - S_{1}(z_{n}) = \pi \epsilon/2 $. 
Next, expanding $S_{1}(\lambda)$ about $z_{n}$, evaluating at $\lambda = z_{n+1}$, and solving for $z_{n+1} - z_{n}$ we obtain
\vspace*{-1ex}
\be
z_{n+1} - z_{n} = \frac{\pi \epsilon}{2|S_{1}'(z_{n})|} + O(\epsilon^{2}), \hspace{2mm} \epsilon \downarrow 0.
\ee
Combining the above results yields
\be
w_{n} + g_{n} =   \frac{\pi \epsilon}{2|S_{1}'(z_{n})|} + O(\epsilon^{2}), \hspace{2mm} \epsilon \downarrow 0.
\label{e:bwgw}
\ee
Finally, Eq.~\eqref{e:wn} and Eq.~\eqref{e:bwgw} together yield Eq.~\eqref{e:WN} for the $n$-th relative bandwidth.

\section{\uppercase{Discussion}}
\label{s:discussion}

In summary, we presented \bb numerical \eb evidence that the semiclassical limit of the focusing NLS equation possesses certain features 
that are relatively independent of the ICs and of whether such ICs are localized or periodic.  
\bb 
Moreover, we tied these numerical observations to an asymptotic characterization of the spectral content of the solutions.  We did so by showing that, \eb 
for a representative class of potentials, the spectrum of the associated scattering problem in the semiclassical limit clusters to the real and imaginary axis of the spectral variable.
This implies that any nonlinear excitations have zero velocity in the semiclassical limit.
We then 
showed that for single-lobe periodic potentials, the spectrum can be analytically characterized using standard asymptotic techniques.
Finally, we computed asymptotic expressions for the relative band width of the nonlinear excitations,
we formulated the concept of effective solitons, and we showed that 
the number of bands scales like $1/\epsilon$ (similarly to the number of discrete eigenvalues for the semiclassical limit on the line \cite{bronski}).
We also showed that, as $\epsilon\downarrow0$, all nonlinear excitations become effective solitons, 
implying that the solution of the focusing NLS equation in the semiclassical limit is described by a 
coherent soliton condensate.

The asymptotic analysis of the spectrum for single-lobe potentials is quite general. 
However, the ICs must be sufficiently ``peaked'' in order for the qualitative features of the temporal evolution in Fig.~\ref{f:1} to arise.
(For example, for ICs with a flat top one can expect behavior such as in~\cite{JM2013,EKT2016}.
See also section~\ref{s:numerics} for another example).\,\ 
At the same time, the properties of the periodic spectrum obtained in \cite{fujiiewittsten2018} are not limited to single-lobe potentials.
Therefore, it is possible that the results of this work apply to a broader class of potentials.
Whether this is indeed the case is an interesting topic for future study.
On the other hand, we strongly emphasize that not all kinds of ICs obviously give rise to the same kind of dynamical behavior
This should not be surprising, since the modulational instability in the focusing NLS equation becomes more and more severe as $\epsilon$ gets smaller, and
the initial-value problem for the associated Whitham modulation equations becomes formally ill-posed in the limit $\epsilon\downarrow0$.
Therefore, one can expect very sensitive dependence of the results with respect to small perturbations,
similarly to what happens in the infinite line~\cite{ClarkeMiller2002}.
Another interesting question is therefore a precise characterization of the ICs that produce the phenomena presented here.

We emphasize that the fact that the behavior in the semiclassical limit is qualitatively the same for localized and periodic ICs 
is limited to the focusing NLS equation.
That is, no such result applies for the KdV equation or the defocusing NLS equation.
This is despite the fact that the WKB analysis is very similar to those for the KdV and defocusing NLS equation in 
\cite{physd2016deng,prl2016deng} and \cite{pre2017deng}, respectively.
The fundamental difference between the defocusing NLS and KdV equations on one hand and the focusing NLS equation on the other hand is that, 
for the former two, each of the the effective solitons produced in the semiclassical limit has a different velocity.
Therefore, these solitons separate from each other, and can be easily identified in the actual solution of the PDE.
In contrast, we showed that for the focusing NLS equation all the bands have zero real part, 
and therefore the effective solitons have zero velocity, 
leading to the formation of a coherent soliton condensate.

\bm
We should note that, physically speaking, the gradient catastrophe is a localized phenomenon, occurring when the compression due to the focusing nonlinearity causes a singularity 
in the dispersionless approximation of the NLS equation,
which is a spatially localized effect.
It is therefore possible the results of \cite{bertolatovbis} may be extended to general cases when a modulated plane undergoes a gradient catastrophe (i.e., a new band is born from the endpoint of the existing band), regardless of the BCs or the behavior of the potential as $x\to\pm\infty$. 
On the other hand, the setting in \cite{bertolatovbis} depends crucially on the BCs
(for example, the fact that the jump in the Riemann-Hilbert problem is confined to the 
real $\zeta$-axis).
Therefore, whether the proofs in \cite{bertolatovbis} easily extend to other settings
remains as an interesting question for further study.
\em

For the ZS problem on the infinite line, there exists a proof that the Lax spectrum of non-negative single-lobe potentials is contained within the real and imaginary axes for all values of $\epsilon>0$ \cite{KS2002,KS2003,BL2018}.
The property does not extend to periodic single-lobe potentials for finite values of $\epsilon$.
The numerical evidence presented in this work, however, suggests that the property applies in the semiclassical limit.

The results of this work open up the obvious problem of characterizing the semiclassical limit in the $xt$-plane.  
Even in the semiclassical limit on the infinite line, a characterization of solutions beyond the secondary breaking curve is still an open problem.
We also emphasize that the genus of the spectral curve arising from the scattering problem in the IST 
(which is independent of $x$ and~$t$)
differs from the genus of the solution in the semiclassical limit, which is local (i.e., dependent on $x$ and~$t$)
and is determined by the semiclassical asymptotics for each fixed value of $x$ and $t$.
For example, for the top right panel of Fig.~\ref{f:1}, both the asymptotics and the numerics of the scattering problem both indicate
a number of spectral bands in excess of 10.
On the other hand, for all $(x,t)$ below the primary caustic, the effective genus of the solution is~0.
It is an interesting open question whether the genus of the spectral curve corresponds to the \textit{maximum} possible value 
of the effective genus in the semiclassical limit.
(For example, in the top right panel of Fig.~\ref{f:1}, only three breakings are visible,
corresponding to a maximum effective genus of 6, which is significantly less than the 10 bands predicted by the spectral problem.
\bb
It is possible that further breakings would appear at later times, but the maximum integration time in the numerical simulations is limited by the
severe growth of round-off error as a result of modulational instability.\eb)

The above is also related to the conjecture, 
formulated in \cite{lyngmiller} for the semiclassical limit on the line,
that an infinite number of caustics arise in the limit $\epsilon\downarrow0$. 
The numerical evolution results shown here suggest that the same conjecture
extends to the problem with periodic BCs.
Indeed, the WKB prediction that the number of bands in the Lax spectrum is $O(1/\epsilon)$
provides a first, indirect, result in support of the conjecture.
On the other hand, to make the WKB rigorous
one should obtain rigorous bounds for the asymptotic approximation of the spectrum obtained with the WKB method.
Doing so is 
outside the scope of this work.

Yet another interesting open question is whether the solutions display recurrence of initial conditions 
(like in the semiclassical limit of the KdV \cite{ZabuskyKruskal} and defocusing NLS equations \cite{pre2017deng}).
It is well known \cite{SatsumaYajima} 
that the evolution of the IC $q(x,0) = \sech x$ with $\epsilon = 1/N$ 
is indeed time-periodic, with temporal period $O(1/\epsilon)$.
More generally, sufficient conditions are also available ensuring the periodicity of 
degenerate solutions of the focusing NLS equation on the line with zero boundary conditions \cite{LiBSchiebold}.
(The term ``degenerate'' indicates solutions produced by purely imaginary discrete eigenvalues.)
Recurrence of ICs has also been shown when few spectral bands are present \cite{lakeyuen,yuenferguson}.
But it is unknown whether recurrence exists for more general single-lobe potentials and generic values of $\epsilon$ 
(either on the line or with periodic ICs).

We expect the results of this work to have broad applicability,
since, similarly to those in \cite{biondinimantzavinos,CPAM2017,biondinilimantzavinos,SIREV}, 
they are almost independent of the details of the initial condition.
Moreover, since the NLS equation arises in many physical contexts, including  
nonlinear optics, deep water waves, acoustics, plasmas and Bose-Einstein condensates,
the results of this work apply to all of these areas.
In particular, nonlinear optical fibers and gravity waves in one-dimensional deep water channels
are especially promising candidates for the experimental verification of the phenomena described here.
Indeed, the phenomena predicted in \cite{biondinimantzavinos,CPAM2017,biondinilimantzavinos,SIREV} 
have recently been observed experimentally in optical fibers \cite{kraych2019}.
We therefore hope that similar settings could provide the vehicle for observing some of the phenomena discussed in this work.

\smallskip
\paragraph{Acknowledgments.}

We thank Percy Deift, Guo Deng, Xudan Luo, Peter Miller, Alex Tovbis and Stefano Trillo for many interesting discussions
\bm
as well as the anonymous reviewers for their thoughtful comments and suggestions.
\em
This work was partially supported by the National Science Foundation under grant numbers DMS-1614623 and DMS-1615524.

\bigskip
\catcode`\@ 11
\def\d@mmyjournal{\errmessage{Reference foul up: nested \journal macros}}
\def\@biblabel#1{#1.}

\end{document}